\documentclass[usenatbib]{mnras}

\usepackage{psfig}
\usepackage{longtable}
\usepackage[dvips]{graphicx}
\usepackage{threeparttable}
\usepackage[T1]{fontenc}
\usepackage{aecompl}

\newcommand{\arcs}{\mbox{\ensuremath{^{\prime\prime}}}}
\newcommand{\arcm}{\mbox{\ensuremath{^{\prime}}}}

\title
[Kinematics of Cygnus~OB2]
{Cygnus OB2 DANCe: A high precision proper motion study of the Cygnus~OB2 association}
\author
[Wright et al.]
{Nicholas J. Wright,$^{1,2}$ Herve Bouy,$^3$ Janet E. Drew,$^2$ Luis Manuel Sarro$^4$\newauthor
Emmanuel Bertin,$^5$ Jean-Charles Cuillandre,$^6$ and David Barrado$^3$ \\
\\
$^{1}$Astrophysics Group, Keele University, Keele, ST5 5BG, UK\\ e-mail: \tt{nick.nwright@gmail.com}\\
$^{2}$Centre for Astrophysics Research, University of Hertfordshire, Hatfield, AL10 9AB, UK\\
$^{3}$Centro de Astrobiolog\'ia, depto de Astrof\'isica, INTA-CSIC, PO BOX 78, 28691, ESAC Campus, \\ 208691 Villanueva de la Ca\~nada, Madrid, Spain\\
$^{4}$Dpt. de Inteligencia Artificial, UNED, Juan del Rosal, 16, 28040 Madrid, Spain\\
$^{5}$Institut d'Astrophysique de Paris, CNRS UMR 7095 and UPMC, 98bis bd Arago, 75014 Paris, France\\
$^{6}$CEA/IRFU/SAp, Laboratoire AIM Paris-Saclay, CNRS/INSU, Universit\'e Paris Diderot, Observatoire de Paris,\\PSL Research University, F-91191 Gif-sur-Yvette Cedex, France\\
}

\begin{document}
\maketitle

\begin{abstract}

We present a high-precision proper motion study of 873 X-ray and spectroscopically selected stars in the massive OB association Cygnus~OB2 as part of the DANCe project. These were calculated from images spanning a 15~year baseline and have typical precisions $< 1$~mas/yr. We calculate the velocity dispersion in the two axes to be $\sigma_\alpha(c) = 13.0^{+0.8}_{-0.7}$ and $\sigma_\delta(c) = 9.1^{+0.5}_{-0.5}$~km~s$^{-1}$, using a 2-component, 2-dimensional model that takes into account the uncertainties on the measurements. This gives a 3-dimensional velocity dispersion of $\sigma_{3D} = 17.8 \pm 0.6$~km~s$^{-1}$ implying a virial mass significantly larger than the observed stellar mass, confirming that the association is gravitationally unbound. The association appears to be dynamically unevolved, as evidenced by considerable kinematic substructure, non-isotropic velocity dispersions and a lack of energy equipartition. The proper motions show no evidence for a global expansion pattern, with approximately the same amount of kinetic energy in expansion as there is in contraction, which argues against the association being an expanded star cluster disrupted by process such as residual gas expulsion or tidal heating. The kinematic substructures, which appear to be close to virial equilibrium and have typical masses of 40--400~M$_\odot$, also do not appear to have been affected by the expulsion of the residual gas. We conclude that Cyg~OB2 was most likely born highly substructured and globally unbound, with the individual subgroups born in (or close to) virial equilibrium, and that the OB association has not experienced significant dynamical evolution since then.

\end{abstract}


\section{Introduction}

Star formation is one of the most important processes in astrophysics, influencing cosmic reionisation, the structure and evolution of galaxies, and the formation of planetary systems. Since most young stars are observed in groups or clusters of some sort \citep[e.g.,][]{carp00,lada03}, understanding the origin of this clustering and the influence it has on the formation and early evolution of stars is critical for a complete theory of star formation. The clustered environment of young stars also affects the formation of planets, through UV photoevaporation from nearby massive stars and close encounters between stars, proto-planetary disks, and young planetary systems \citep{adam06}.

Star clusters are often considered a fundamental unit of star formation \citep{pfla09,pfal09}, with star formation occurring in quantised, relatively dense and gravitationally-bound systems that form embedded within molecular clouds. It has been suggested that the vast majority of stars form within these clusters \citep[e.g.,][]{carp00,krum14}, and that the dense clustering of protostars may play an important role in how stars build up their masses \citep[e.g.,][]{zinn82,bonn01}.

While the majority of stars are observed to be clustered at a young age, only $\sim$10\% of stars are found in bound clusters by 10~Myr \citep{lada03}. The most common explanation for the supposed disruption of these clusters is that the feedback-driven dispersal of the residual gas left over from star formation lowers the gravitational potential holding the cluster in virial equilibrium, leaving the cluster in a super-virial state and prone to expansion and dispersion \citep[e.g.,][]{hill80,lada84,good06,baum07}. Another possible explanation is that a gravitationally bound star cluster could be disrupted by tidal heating from the surrounding giant molecular clouds (GMCs) in the region it was born \citep[e.g.,][]{elme10,krui11b}. Regardless of the mechanism this expanded state would be briefly visible as a low density group of young stars known as an {\it association} \citep[e.g.,][]{blaa64,brow97,krou01}, before dispersing into the Galactic field.

An alternative view of star formation suggests that stars form over a wide range of initial densities from loose groups up to dense clusters in a hierarchical structure that originates from the structure of the parental molecular cloud \citep[e.g.,][]{elme02,bast07,elme08,bonn11}. This picture explains the presence of young stars over a wide range of densities \citep{bres10}, including both isolated young stars of any mass \citep{lamb10} and low-density associations \citep{wrig14b}. In this scenario the densest groups collapse to form bound and long-lived star clusters while the low density groups naturally disperse as associations without passing through a densely clustered phase \citep[e.g.,][]{elme01,krui12}. The origin of associations, of both OB or T varieties, therefore provides a valuable discriminant to distinguish between different models of star formation.

The kinematics of young stars can provide powerful constraints on theories of star formation, particularly if radial velocities (RVs) and proper motions (PMs) are combined to construct true 3-dimensional space velocities for large numbers of stars. Recent, large-scale spectroscopic surveys such as the {\it Gaia}-ESO\footnote{The European Southern Observatory.} Survey or IN-SYNC\footnote{The INfrared Spectra of Young Nebulous Clusters program.} are beginning to provide RVs for large numbers of stars in nearby star forming regions and clusters \citep[e.g.,][]{jeff14,fost15}, but transverse PM velocities are currently lacking. {\it Gaia} \citep{perr01} will ultimately provide PMs for a billion stars in our Galaxy down to $\sim$20$^{th}$ magnitude, though a full data release is not expected before 2022.

High-precision astrometry and PMs can now be extracted from well-calibrated ground-based, wide-field exposures if sufficient, high-quality data are available. This is the goal of the DANCe (Dynamical Analysis of Nearby Clusters) survey program \citep{bouy13}. In this paper we adopt this method to calculate PMs for stars in the massive OB association Cyg~OB2 to study its kinematics and attempt to constrain its origin. Cyg~OB2 is one of the most massive OB associations in our Galaxy with a total stellar mass of $\sim$1--3~$\times 10^4$~M$_\odot$ \citep{drew08,wrig10a,wrig15a} and home to many hundreds of massive stars with masses up to $\sim$100~M$_\odot$ \citep[e.g.,][]{mass91,come02,hans03,kimi07,wrig15a}. Furthermore at a distance of only 1.33~kpc \citep{kimi15}\footnote{Throughout this work we will use the eclipsing binary distance of $1.33 \pm 0.06$~kpc calculated by \citet{kimi15}, in good agreement with the parallax distance of $1.40 \pm 0.08$~kpc calculated by \citet{rygl12} for parts of the surrounding Cygnus~X GMC.} it can be studied in sufficient detail to resolve and characterise the kinematics of both high- and low-mass stars. The majority of the stars in Cyg~OB2 do not have RVs available for them \citep[with the exception of 120 OB stars for which moderate precision, $\sigma_{RV} \sim 5$--10~km/s, RVs are available,][]{kimi07,kimi08b}, and therefore this paper represents the first large kinematic study of the region.

This paper is outlined as follows. In Section~2 we present the observations used to derive our PMs, outline the method used, and discuss the selection of Cyg~OB2 members from within our PM catalogue. In Section~3 we present the 2-dimensional PM velocity distributions and calculate the velocity dispersions. In Section~4 we present and discuss the PM vector map, and study the evidence for contraction, expansion, rotation, and kinematic substructure. In Section~5 we discuss the implications of our results for our understanding of the formation and evolution of Cyg~OB2 and of OB associations in general.

\section{Observations}

\subsection{Data used}

Table~\ref{instruments} summarises the observations, instruments, and telescopes used in this work. To produce the most accurate PMs we searched the public archives of observatories from around the world for wide-field images within 1$^\circ$ of the centre of Cyg~OB2, which is commonly regarded as Cyg~OB2 \#8, the trapezium of O stars at RA 20:33:16, Dec +41:18:45 \citep[e.g.,][]{schu56,vink08}. The archival data that were gathered included 2631 different observations.

\begin{table*}
\caption{Instruments used in this study. $N_{obs}$ refers to the number of separate exposures within each observational dataset.} 
\label{instruments}
\begin{tabular}{llccccccc}
\hline
Year &	Observatory & Instrument & Filters 					& Platescale & Chip layout & Chip size & Field of view & $N_{obs}$ \\
 & & & & [pixel$^{-1}$] & & & [deg$^2$] & (by filter)\\
 \hline
1998	 & JKT$^1$	& SITe2 CCD$^2$	& H$\alpha$			& 0.333\arcs\	& $1 \times 1$	& 2k $\times$ 2k & $10\arcm \times 10\arcm$ & 43 \\ 
1999 & Kiso 1.05m$^3$ & 2k CCD$^4$	& $R$		& 1.5\arcs\ & $1 \times 1$ & 2k $\times$ 2k & $51\arcm \times 51\arcm$ & 9\\ 
2003 & APO 2.5m$^5$ & SDSS$^6$ 		& $u$, $g$, $r$, $i$, $z$	& 0.4\arcs\	& $6 \times 1$ & 2k $\times$ 2k & $82\arcm \times 14\arcm$ & 132 of each \\ 
2003 & INT$^7$& WFC$^8$	& $B$, $V$			& 0.333\arcs\	& $3 \times 1$ $+1$ & 2k $\times$ 4k & $34\arcm \times 34\arcm$ & 2, 3\\ 
2003 & INT	& WFC	& $r$, $i$, H$\alpha$	& 0.333\arcs\	& $3 \times 1$ $+1$ & 2k $\times$ 4k & $34\arcm \times 34\arcm$ & 125, 174, 128\\
2004 & INT	& WFC	& $B$, $V$			& 0.333\arcs\	& $3 \times 1$ $+1$ & 2k $\times$ 4k & $34\arcm \times 34\arcm$ & 1, 2\\ 
2004 & INT	& WFC	& $r$, $i$, H$\alpha$	& 0.333\arcs\	& $3 \times 1$ $+1$ & 2k $\times$ 4k & $34\arcm \times 34\arcm$ & 88, 113, 89\\
2005 & INT	& WFC	& $r$, $i$, H$\alpha$	& 0.333\arcs\	& $3 \times 1$ $+1$ & 2k $\times$ 4k & $34\arcm \times 34\arcm$ & 4, 8, 3\\
2006 & INT	& WFC	& $U$, $g$, $r$, $i$		& 0.333\arcs\	& $3 \times 1$ $+1$ & 2k $\times$ 4k & $34\arcm \times 34\arcm$ & 31, 31, 15, 3\\
2006 & UKIRT$^9$ & WFCAM$^10$ & $J$, $H$, $K$		& 0.4\arcs\	& $2 \times 2$	& 2k $\times$ 2k & $40\arcm \times 40\arcm$ & 46, 48, 80\\
2007 & INT	& WFC	& $U$, $g$, $r$, $i$, H$\alpha$	& 0.333\arcs\	& $3 \times 1$ $+1$ & 2k $\times$ 4k & $34\arcm \times 34\arcm$ & 35, 48, 59, 14, 6\\ 
2007 & UKIRT & WFCAM& $J$					& 0.4\arcs\	& $2 \times 2$	& 2k $\times$ 2k & $40\arcm \times 40\arcm$ & 8\\
2008 & UKIRT & WFCAM& $J$, $H$, $K$		& 0.4\arcs\	& $2 \times 2$	& 2k $\times$ 2k & $40\arcm \times 40\arcm$ & 48 of each\\
2008 & CFHT & WIRCam$^{11}$ & $K_s$, $Br\gamma$, H$_2$	& 0.306\arcs\ & $2 \times 2$ & 2k $\times$ 2k & $22\arcm \times 22\arcm$ & 106, 62, 45\\ 
2009 & GTC$^{12}$	& OSIRIS$^{13}$	& $i$, $r$, $z$			& 0.127\arcs\ & $2 \times 1$ & 2k $\times$ 4k & $8\arcm \times 8\arcm$ & 40, 44, 45\\ 
2011 & UKIRT & WFCAM & $K$			& 0.4\arcs\	& $2 \times 2$	& 2k $\times$ 2k & $40\arcm \times 40\arcm$ & 80\\
2011 & GTC	& OSIRIS	& $i$, $r$, $z$			& 0.127\arcs\ & $2 \times 1$ & 2k $\times$ 4k & $8\arcm \times 8\arcm$ & 53, 58, 54\\ 
2011 & Calar Alto 3.5m & Omega 2000$^{14}$	 & $K_s$			& 0.45\arcs\ & $1 \times 1$ & 2k $\times$ 2k & $15\arcm \times 15\arcm$ & 111\\ 
2012 & KPNO 4m$^{15}$	& Mosaic 1$^{16}$	& WRC4 5825\AA\		& 0.26\arcs\ & $4 \times 2$ & 2k $\times$ 4k & $36\arcm \times 36\arcm$ & 24 \\ 
2012 & CFHT	& MegaCam$^{17}$	& $u$, $g$, $r$, $i$	& 0.187\arcs\ & $4 \times 9$ & 2k $\times$ 4k & $58\arcm \times 57\arcm$ & 16 of each\\ 
2013 & CFHT	& MegaCam	& $u$, $g$, $r$, $i$		& 0.187\arcs\ & $4 \times 9$ & 2k $\times$ 4k & $58\arcm \times 57\arcm$ & 5, 5, 9, 60\\ 
\hline
\end{tabular} 
\begin{tablenotes}[para,flushleft]
Notes and references: 
$^1$ The Jacobus Kapteyn Telescope. 
$^2$ Website: http://www.ing.iac.es/Astronomy/observing/manuals/ps/jkt\_instr/jag.pdf. 
$^3$ Kiso Observatory, University of Tokyo, Japan. 
$^4$ Website: http://www.ioa.s.u-tokyo.ac.jp/kisohp/INSTRUMENTS/instruments\_e.html. 
$^5$ Apache Point Observatory 2.5m telescope. 
$^6$ The Sloan Digital Sky Survey \citep{york00}. 
$^7$ The Isaac Newton Telescope. 
$^8$ The Wide Field Camera \citep{ives98}.
$^9$ The United Kingdom Infrared Telescope. 
$^{10}$ The Wide-Field CAMera \citep{casa07}. 
$^{11}$ The Wide-Field Infrared Camera \citep{puge04}. 
$^{12}$ Gran Telescopio Canarias. 
$^{13}$ Optical System for Imaging and low Resolution Integrated Spectroscopy, obtained as part of observations by \citet{guar12}. 
$^{14}$ \citet{bail00}. 
$^{15}$ The Kitt Peak National Observatory Mayall Telescope. 
$^{16}$ \citet{wolf00}. 
$^{17}$ \citet{boul03}. 
\end{tablenotes}
\end{table*}

To improve the PMs by extending the time baseline and number of epochs we complemented the archival data by obtaining new deep wide-field observations of Cyg~OB2. These observations were obtained with the Omega 2000 camera on the Calar Alto 3.5m telescope in 2011 and the MegaCam instrument on the Canada France Hawaii Telescope (CFHT) during 2012 and 2013. These observations were all designed to optimise the astrometric calibration and consisted of multiple pointings covering a $\sim$1$^\circ$ area centred on Cyg~OB2, with each exposure dithered and offset by $\sim$2--4$^{\prime}$ in RA and Dec. The overlap between observations ensures an accurate astrometric anchoring over the entire survey area. All the observations included exposures taken in the $i$ or $K_s$-bands where differential chromatic refraction (DCR) is lower and the seeing is often better. These observations brought the total number of observations obtained for this work to 2885, from a total of 9 observatories and 10 different instruments.

The quality of the PMs calculated in this work is dependent on the positional accuracy achieved in individual epochs and is therefore influenced by the signal-to-noise ratio of individual stellar measurements, the full width at half maximum (FWHM) of the point sources, the sampling (pixel scale) of the images, and the airmass of the observations (because atmospheric turbulence and differential chromatic refraction quickly increase with airmass). The majority of observations were obtained at airmass $< 1.2$ with the 90th percentile at airmass 1.29. The median airmass is 1.07.

The individual raw images were processed using an updated version of {\sc Alambic} \citep{vand02}, a software suite developed and optimised for the processing of data from large multi-CCD imagers and adapted for the instruments used here. {\sc Alambic} includes standard processing procedures such as overscan and bias subtraction, flat-field correction, bad-pixel masking, chip-to-chip gain harmonisation (for multi-chip cameras), de-stripping and fringing correction (when needed), and non-linear correction (for infrared detectors). All these steps are performed independently on each read-out port whenever several ports are present.

\subsection{Astrometric analysis}

\begin{figure}
\begin{center}
\includegraphics[height=240pt, angle=270]{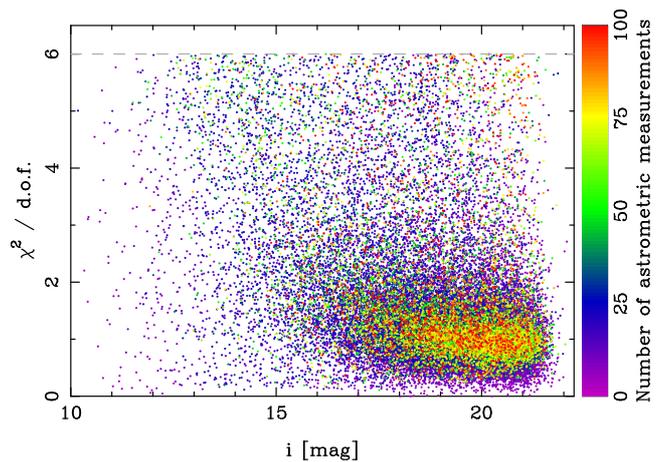}
\caption{Reduced $\chi^2$ of the PM fits as a function of the INT $i$-band magnitude (where available), with the number of astrometric measurements used in the PM fit indicated by the colour. The cut-off at $\chi^2 / \mathrm{d.o.f.} = 6$ corresponds to the outlier rejection threshold. For clarity only 10\% of the catalogue is shown. The majority of PMs for Cyg~OB2 sources have baselines $\sim$15~years and therefore PM uncertainties $< 1$~mas/yr.}
\label{chi2}
\end{center}
\end{figure}

Astrometry was performed with the Astr$O$matic\footnote{http://www.astromatic.net} software suite, including {\sc SExtractor} \citep[Source Extractor,][]{bert96}, SCAMP \citep[Software for Calibrating AstroMetry and Photometry,][]{bert06}, and PSFEx \citep[Point Spread Function Extractor,][]{bert11}. The whole process is described in detail in \citet[][see Section~7]{bouy13}, but we briefly outline the most important steps here:

\begin{enumerate}
\item Recover and equalise image metadata. Many astrometric tasks require parameters specific to each observatory, instrument, or observation. These were gathered and brought onto the same FITS metadata standard.
\item Modelling the PSF. An accurate model of the PSF is needed for every exposure of every chip from every instrument, which sometimes must be performed at the sub-pixel level if the images are significantly under-sampled (such as those with good seeing). This was performed in a non-parametric way with the PSFEx software.
\item Cataloguing. For sources with more than three pixels above 1.5 standard deviations of the local background {\sc SExtractor} was used to measure fluxes and positions using the empirical PSF. In contrast to iterative Gaussian centroiding, PSF model fitting is mostly immune to spatial discretisation effects caused by under-sampling, and also allows saturated pixels to be censored without excessively degrading the positional accuracy of (moderately) saturated stars.
\item Quality assurance. Not all archive data are of sufficient quality to produce reliable and accurate astrometric measurements. All exposures were screened for defects using both semi-automated quality-control based on PSFEx and {\sc SExtractor} measurements, and manual inspection of astrometric measurements in different exposures as a function of the different instruments used, the observing conditions (e.g., airmass), and properties such as extraction flags and measured magnitudes. Astrometric measurements that were flagged by {\sc SExtractor} as saturated or with truncated PSFs (close to an image boundary) were rejected. {\sc SExtractor}'s PSF fitting module reduces the impact of this \citep[see e.g.,][]{bouy13}, though minor magnitude-dependent astrometric biases were noted for some saturated stars.
\item Estimating astrometric uncertainties. Positional uncertainties are important when calculating the global astrometric solution and for computing the weightings needed for calculating PMs. Our estimated positional uncertainties take into account photon noise, relative motions caused by atmospheric turbulence, and imperfect deblending of close sources \citep[see Section~7.5 of][]{bouy13}. 
\item Computing global astrometric solutions. This is computed iteratively by SCAMP by minimising the quadratic sum of differences in position between overlapping detections from pairs of catalogues. This requires the calculation of a {\it reprojection operator} for each ``astrometric instrument'' (defined as the unique combination of camera, filter, and observing run, and distinct from the traditional meaning of the word instrument referring to the camera used on the telescope itself). This must be calculated for each observing run because chip distortion patterns can change from run to run as instruments are often taken off telescopes between runs. Based on header information and logbooks we have identified 113 different astrometric instruments, taken with 10 different traditional instruments through 29 different filters.
\item Fitting individual PMs. After the second iteration of global astrometric calibration, moving sources are identified by cross-matching different observations in time order with a cross-matching radius of 3$^{\prime\prime}$. Once cross-matched, PMs are calculated by SCAMP using a weighted linear fit to source positions as a function of time. No attempt was made to include the effect of trigonometric parallax in the fit because at a distance of 1.33~kpc the maximum amplitude of the parallax motion is only $\sim$0.7~mas/yr (the effect will be stronger for nearby stars, which could influence our global astrometric solution, though their numbers are likely to be in the minority given the range of photometric magnitudes we are sensitive to). To filter out poor astrometric data, any PM fit with a reduced $\chi^2 / \mathrm{d.o.f.} > 6$ are re-calculated after removing the astrometric measurement that has increased $\chi^2 / \mathrm{d.o.f.}$ the most (this threshold was chosen to balance rejecting too many measurements and keeping very poor fits). This process is repeated until either $\chi^2 / \mathrm{d.o.f.} \leq 6$ or a maximum of 20\% of points have been removed \citep[see][for more details for this process]{bouy13}. The resulting $\chi^2 / \mathrm{d.o.f.}$ values for the fits are shown in Figure~\ref{chi2}.
\end{enumerate}

\begin{figure*}
\begin{center}
\includegraphics[height=500pt, angle=270]{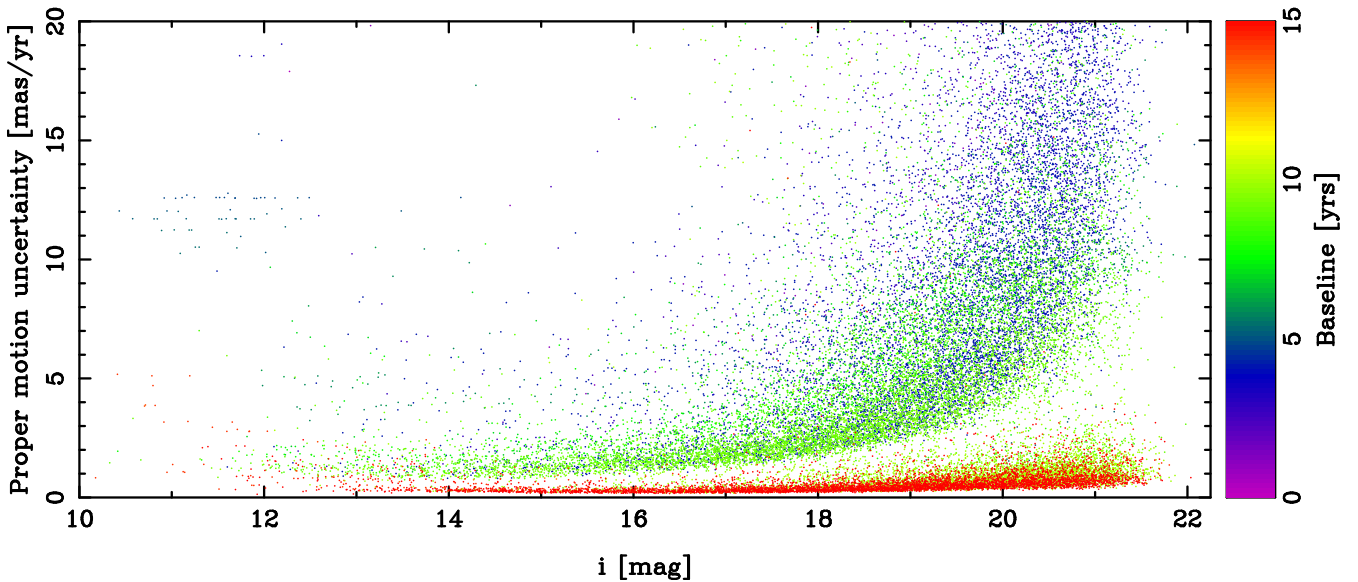}
\caption{PM uncertainty as a function of INT $i$-band magnitude (where available) with the baseline of the astrometric measurements used shown by the colour. For clarity only 10\% of the catalogue is shown. While Gaia will obtain more precise PMs for stars with $i < 18$~mag, stars fainter than this will not be detected by Gaia.}
\label{errors}
\end{center}
\end{figure*}

The positions and PMs calculated by this method are not tied to an absolute reference system such as the International Celestial Reference System (ICRS) but are instead calculated relative to each other (a method that is valid for separations up to a few degrees, beyond which distance-dependent ``drifting'' biases become non-negligible). Our measurements could be placed on the ICRS by comparing them to an existing astrometric catalogue on the ICRS, however most stars from the {\sc Hipparcos} and Tycho catalogues in our field of view are bright and hence saturated, precluding their use. An alternative method is to use the PMs of extragalactic sources in our field of view and determine the offset required to reduce their PMs to zero. \citet{bouy13} found that this method added a $\sim$0.3~mas/yr uncertainty to the individual relative PMs. Since our scientific objectives only require relative PMs we have decided not to tie our PMs to the ICRS to avoid this increased uncertainty. This decision could risk introducing a small PM gradient over the field produced by the Milky Way disk, but since our field of view is small it will be a very minor effect. The fact that we do not observe any correlation between our PMs and Galactic latitude suggests that this hasn't seriously biased our measurements.

\subsection{Astrometric accuracy and its limitations}
\label{s-uncertainties}

Astrometric accuracy is mainly limited by the distortion correction and the variability that arises from atmospheric turbulence. Further noise is added to the astrometric solution by cosmic rays, bad pixels, artefacts produced by saturated stars, and chromatic centroid shifts from extragalactic sources, nebulae and unresolved multiple systems. The contribution of all of these issues can be greatly minimised by selecting only point-like sources and rejecting outliers in the PM fits. DCR, the wavelength-dependent shift of the centroid due to dispersive elements along the path, can also affect the astrometry. However, since the vast majority (92\%) of our observations were obtained at airmass $< 1.4$ (for which DCR offsets are typically low) and in the red or near-infrared part of the spectrum (where the amplitude of the DCR is also smaller), this is not expected to be a significant source of uncertainty.

Figure~\ref{errors} shows a representative sample of the estimated PM uncertainty as a function of $i$-band magnitude, with the points colour-coded based on the total baseline from which the PMs are calculated. The PM uncertainty is dependent on both the magnitude and the baseline of the observations, with lower uncertainties for brighter sources and longer baselines. For those sources with baselines $\sim$15~years (including the majority of sources we are interested in), the PM uncertainty is typically $<$1~mas/yr and $\sim$0.4~mas/yr for sources brighter than $i \sim 18$~mag (equivalent to stellar masses of 1--1.5~M$_\odot$ at the distance and extinction of Cyg~OB2).

While Gaia \citep{debr12} will obtain more precise PMs for stars with $i < 18$~mag (based on Gaia's limiting magnitude of $G \sim 20$~mag and the typical $G - i$ colours of these stars of $\sim$2~mag) it will not detect any of the stars fainter than this, which will only be improved in the foreseeable future by PMs from the Large Synoptic Survey Telescope \citep{ivez08}.

\begin{figure*}
\begin{center}
\includegraphics[height=400pt, angle=270]{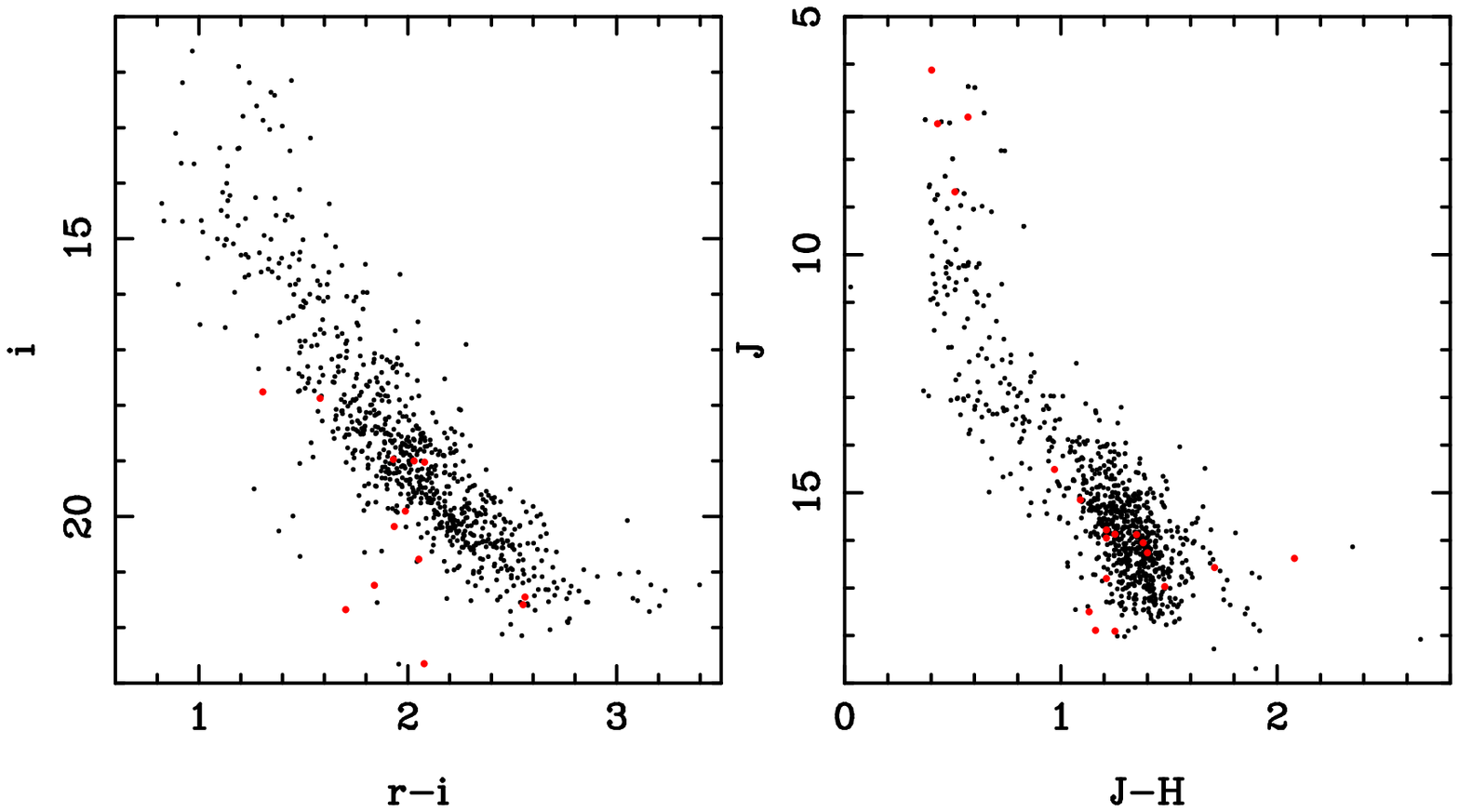}
\caption{Colour-magnitude diagrams illustrating the relative completeness of our PM sample based on the X-ray catalogue of members from \citet{wrig09a} and OB star catalogue of \citet{wrig15a}. Members of Cyg~OB2 with measured PMs are shown as black dots and those without PMs are shown as red dots. {\it Left:} Optical $i$ vs $r-i$ colour-magnitude diagram using IPHAS \citep{drew05} and OSIRIS \citep[][converted onto the IPHAS photometric system]{guar12} photometry for 771 of the 873 sources with PMs and 13 of the 19 sources without PMs. {\it Right:} Near-IR $J$ vs $J-H$ colour-magnitude diagram using 2MASS \citep{skru06} and UKIDSS \citep{luca08} photometry for 851 of the 873 sources with PMs and all 19 sources without PMs. All the sources with or without PMs from both \citet{wrig09a} and \citet{wrig15a} are shown in one of the two panels.}
\label{completeness}
\end{center}
\end{figure*}

\subsection{Cygnus OB2 membership selection}

The objective of this work is to produce a sample of Cyg~OB2 members with measured PMs free from kinematic biases. For this reason we have not used the kinematic measurements to identify new members of the OB association, as this can introduce biases into the kinematics, but instead base our membership selection on purely non-kinematic criteria.

To identify the low-mass members of Cyg~OB2 we cross-matched our PM catalogue with the X-ray source list of \citet{wrig09a}. X-ray observations provide a largely unbiased diagnostic of youth that is highly effective in separating young association members from older field stars because pre-main-sequence stars are typically 10-1000 times more luminous in X-rays than main-sequence stars \citep{prei05}. This is because young stars rotate much more rapidly than older field stars and thus through the actions of the magnetic dynamo (which operates in stars with radiative cores and convective envelopes) have higher levels of magnetic activity that are manifest through enhanced X-ray emission \citep[e.g.,][]{wrig11b}. The only exception to this are A- and late-B-type stars that do not appear to emit X-rays, most likely due to the lack of a convective envelope \citep{schm97}.

The deeper of the two X-ray observations studied by \citet{wrig09a} is centred on the core of the association and is estimated to be complete and spatially unbiased in the mass ranges of 0.8--1.7~$M_\odot$ and $> 5$~M$_\odot$ \citep{wrig14b}. \citet{wrig10a} studied the properties of these sources in detail, using optical and near-IR photometry to identify and remove foreground contaminants and estimate stellar masses, and it is this source list that we use here. We extend the decontamination process by selecting only sources that were also detected by the Cygnus~OB2 {\it Chandra} Legacy Survey \citep{wrig14c}, excluding $\sim$150 faint sources that did not pass the more conservative source verification process used by \citet{wrig14c} and leaving 867 sources. Of these, 848 were successfully cross-matched (using a matching radius of 1$^{\prime\prime}$) to our PM catalogue, a success rate of 98\% (leaving 19 X-ray sources without PMs).

We also cross-matched our PM catalogue with the census of OB stars in Cyg~OB2 compiled by \citet{wrig15a}, for which stellar masses and ages were calculated from rotating stellar evolutionary models. Limiting ourselves to the 58 OB stars that fall within the field of view of the X-ray observations and using the same cross-matching radius, we obtained 56 matches, 20 O-type stars and 36 B-type stars.

Combining the X-ray and OB star samples, and removing duplications from the two catalogues, gives a total of 873 members of Cyg~OB2 with measured PMs (between the two catalogues only 19 sources do not have PMs). Figure~\ref{completeness} shows the distribution of known members of Cyg~OB2 in both optical and near-IR colour-magnitude diagrams \citep[additional optical photometry taken from][]{guar12}, with the sources without PMs shown in red. The small fraction of sources without PMs includes some of the brightest sources in the near-IR colour-magnitude diagram (for which PMs can be difficult to calculate due to saturation) as well as some of the faintest (which may not be detected in enough observations to yield accurate PMs).

\begin{figure}
\begin{center}
\includegraphics[height=220pt, angle=270]{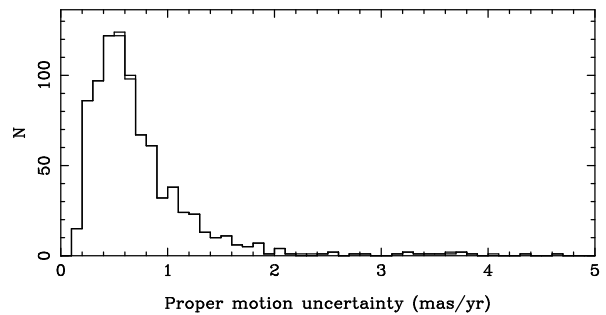}
\caption{PM uncertainty distributions along both axes for 864 of the 873 stars in our sample. Of the 9 stars not shown on this figure, all have PM uncertainties $<$~12~mas/yr in both dimensions.}
\label{uncertainties}
\end{center}
\end{figure}

For this sample the median number of astrometric measurements used for the PMs is 77, with 10th and 90th percentiles of 43 and 107. Figure~\ref{uncertainties} shows the PM uncertainty distribution, in both axes for our sample \citep[almost identical because the astrometric precision is the same in both axes, see][]{bouy13}. The vast majority of stars ($\sim$80\%) have estimated PM uncertainties $<$~1~mas/yr, with a median of 0.59~mas/yr in both dimensions (equivalent to 3.7~km/s at a distance of 1.33~kpc), and a small tail of stars with high uncertainties up to $\sigma = 12$~mas/yr. Since we have calculated relative PMs we set the medians of our PM distributions to be zero in both dimensions by applying small non-zero offsets to the PMs. The full catalogue is provided in Table~\ref{catalogue}.

\begin{table*}
\caption{Proper motions for members of Cyg~OB2.}
\label{catalogue} 
\begin{tabular}{lccccccccc}
\hline
DANCe ID		& R.A.	& Dec.	& $\mu_\alpha$	& $\sigma_{\mu_\alpha}$	& $\mu_\delta$	& $\sigma_{\mu_\delta}$	& ID$_\mathrm{X}$ & \multicolumn{2}{c}{ID$_{\mathrm{OB}}$} \\
\cline{9-10}
					& (J2000)		& (J2000)		& (mas/yr)		& (mas/yr)		& (mas/yr)		& (mas/yr)		& 		& MT91 & S58 \\
\hline
J203227.69+411316.9	& 20:32:27.69 	& +41:13:16.9	& -0.78		& 0.50		& -0.80		& 0.50		& 341       \\
J203229.12+411401.3	& 20:32:29.12 	& +41:14:01.3	& -1.15		& 1.03		& -1.23		& 1.03		& 355       \\
J203229.29+410849.4	& 20:32:29.29 	& +41:08:49.4	& +0.22		& 0.21		& +0.05		& 0.21		& 357       \\
J203229.90+411453.3	& 20:32:29.90 	& +41:14:53.3	& +0.78		& 2.07		& -2.02		& 2.08		& 363       \\
J203230.39+411006.5	& 20:32:30.39 	& +41:10:06.5	& +1.40		& 1.24		& -1.62		& 1.24		& 366       \\
J203230.64+410856.4	& 20:32:30.64 	& +41:08:56.4	& -2.02		& 0.75		& -1.12		& 0.75		& 371       \\
J203230.66+410831.0	& 20:32:30.66 	& +41:08:31.0	& -1.66		& 0.94		& +0.71		& 0.94		& 369       \\
J203231.44+410955.8	& 20:32:31.44 	& +41:09:55.8	& -0.84		& 0.65		& +0.09		& 0.65		& 379       \\
J203231.54+411408.3	& 20:32:31.54 	& +41:14:08.3	& +2.76		& 0.34		& -0.04		& 0.34		& 383	& 267   \\
J203232.49+411313.2	& 20:32:32.49 	& +41:13:13.2	& -0.26		& 0.34		& -1.20		& 0.34		& 391       \\
\hline
\end{tabular}
\newline \flushleft Notes. ID$_\mathrm{X}$ refers to the source number from the X-ray catalogue of \citet{wrig09a}. ID$_{\mathrm{OB}}$ lists two source identification numbers typically used for the OB stars in Cyg~OB2 \citep[see Table~B1 of][for details]{wrig15a}.\\
(This table is available in its entirety in a machine-readable form in the online journal. A portion is shown here for guidance regarding its form and content.)
\end{table*}

\section{Proper motion velocity dispersions}

\begin{figure*}
\begin{center}
\includegraphics[height=500pt, angle=270]{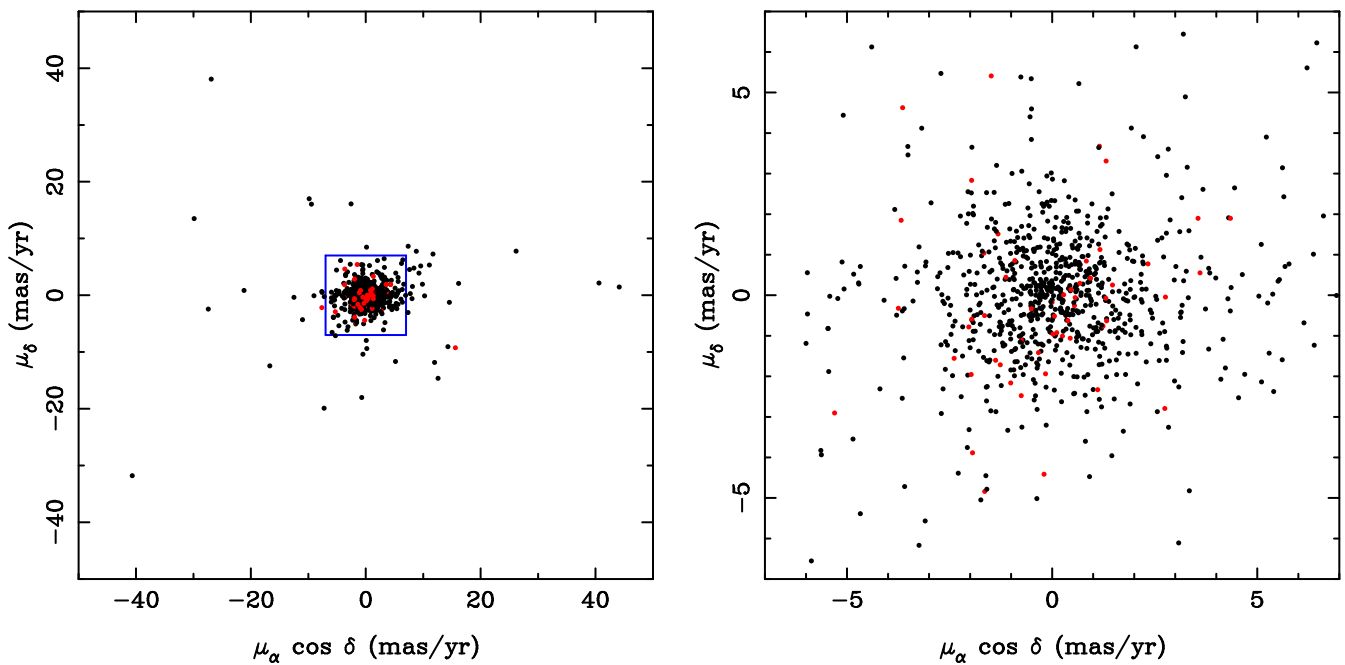}
\caption{Vector point diagram for members of Cyg~OB2 with PMs (note that PMs are not placed on an absolute reference system but are only calculated {\it relative} to each other) with the spectroscopically-known OB stars shown as red dots. The left-hand panel shows 869 of the 873 stars with PMs (four stars with very high proper motions are not shown), highlighting the presence of a number of runaways or possible contaminants with high PMs. The right-hand panel (inset shown in the left-hand panel with a blue box) shows a close-up of the centre of the velocity space, with 822 of the 873 stars shown.}
\label{vector_point}
\end{center}
\end{figure*}

In this section we use the PMs for members of Cyg~OB2 to calculate the 2-dimensional velocity dispersion and assess the dynamical state of the association. The velocity dispersion can provide information on the boundedness of a group of stars, the dynamical conditions and the extent of dynamical evolution within the region. PMs provide a more accurate measure of the intrinsic velocity dispersion of a group of stars than RVs because they are not noticeably affected by the motions of individual stars within binary systems (because PMs provide a measure of the motions of stars integrated over a baseline and not an instantaneous measure of the velocity as RVs do). In the text that follows we denote the velocity dispersions in RA and Dec as $\sigma_\alpha$ and $\sigma_\delta$, though technically the velocity dispersions of the PMs in RA and Dec should be written as $\sigma_{\mu_\alpha \, \mathrm{cos} \, \delta}$ and $\sigma_{\mu_\delta}$.

Figure~\ref{vector_point} shows the vector point diagram of PMs for the majority of Cyg~OB2 members. The PMs of non-members of Cyg~OB2 are not shown, but these have a similar distribution, implying that Cyg~OB2 members cannot be kinematically separated from the field population. A number of outliers with very large relative PMs are evident in this diagram and are discussed in more detail in Section~\ref{s-outliers}. These outliers cause the velocity dispersion (Section~\ref{s-veldisp}) to come out very differently for a Gaussian dispersion compared to that calculated from an outlier-resistant method such as the interquartile range (IQR), and the use of a forward modelling approach (Section~\ref{s-model}) reinforces this.

\subsection{Calculating the velocity dispersions}
\label{s-veldisp}

\begin{figure}
\begin{center}
\includegraphics[height=240pt, angle=270]{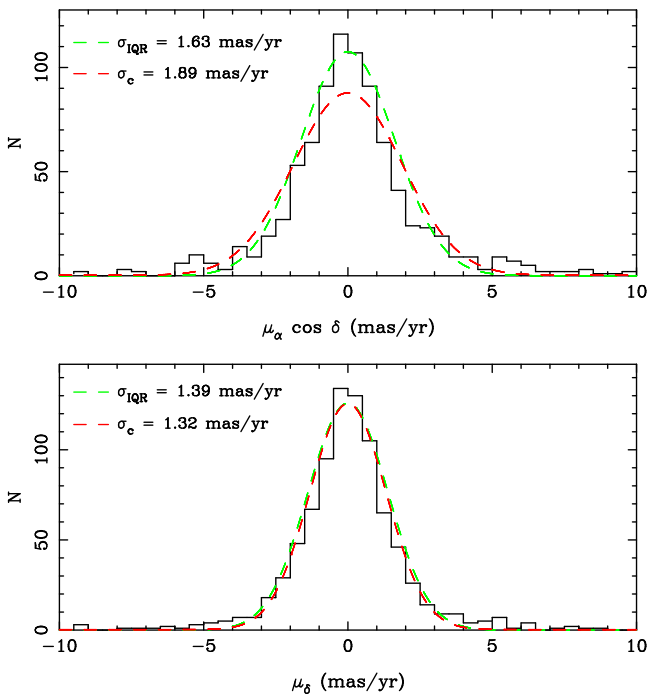}
\caption{PM velocity distributions along both axes (black histogram) showing the interquartile range velocity dispersions (green dashed lines) and the two-component Gaussian mixture model for the velocity dispersion (red dashed lines). The best fitting velocity dispersions are listed in each case, which for the Gaussian mixture model is the dominant Cyg~OB2 component.}
\label{velocity_dispersions}
\end{center}
\end{figure}

Figure~\ref{velocity_dispersions} shows the binned velocity distribution (over the range covered by the majority of stars), which has a broadly Gaussian distribution with low-amplitude, broad wings. The simplest estimate of the velocity dispersion is the standard deviation, which for our sample of 873 stars gives $\sigma_\alpha(std) = 8.54^{+0.20}_{-0.03}$~mas/yr and $\sigma_\delta(std) = 7.35^{+0.18}_{-0.01}$~mas/yr (uncertainties calculated from a Monte Carlo simulation in which the individual PMs are varied by their estimated uncertainties). These velocity dispersions are considerably larger than the observed distribution because they are being enlarged by a number of outliers with large velocity offsets. There are 23 (17) stars with $|\mu| > 10$~mas/yr in the RA (Dec) dimension, and 1 (2) of these have $|\mu| > 100$~mas/yr. These outliers could be removed by a process of ``sigma clipping'', but this is a procedure fraught with difficulties of subjectivity and irreproducibility.

An alternative estimate of the velocity dispersion can be derived using an outlier-resistant method that assumes that the underlying dispersion is approximately Gaussian (as it does appear to be, see Figure~\ref{velocity_dispersions}), but with broad, low-amplitude wings. One of the most common such diagnostics is the interquartile range (IQR $= q_{75} - q_{25}$), which is related to the velocity dispersion by

\begin{equation}
\sigma(IQR) = 0.741 \, (q_{75} - q_{25})
\end{equation}

\noindent where $q_{25}$ and $q_{75}$ are the 25th and 75th percentiles of the velocity distributions and the normalising factor of 0.741 is the reciprocal of the interquartile range of a Gaussian distribution with a standard deviation of one. This method gives velocity dispersions of $\sigma_\alpha(IQR) = 1.63^{+0.27}_{-0.17}$ and $\sigma_\delta(IQR) = 1.39^{+0.22}_{-0.12}$~mas/yr, which provide a much better fit to the observed velocity distributions (Figure~\ref{velocity_dispersions}). The true velocity dispersion will be less than this, having been broadened by the not-insignificant (and highly heteroskedastic) uncertainties for our PM measurements. A first-order approximation of the underlying velocity dispersion can be obtained by subtracting the square of the median uncertainty (on the assumption that while the uncertainties are non-uniform they do not vary significantly), which gives $\sigma_\alpha(IQR_0) = 1.5 \pm 0.3$ and $\sigma_\delta(IQR_0) = 1.3 \pm 0.2$~mas/yr.

\subsection{Forward modelling the velocity dispersions}
\label{s-model}

An alternative method to calculate the underlying velocity dispersion is to construct a model that takes into account both the velocity dispersion of Cyg~OB2 members, the distribution of kinematic outliers, and the individual measurement uncertainties, and then find the model parameters that best reproduce the observations. We implement this by modelling the velocity distribution using two Gaussians in each dimension, each with their own velocity dispersions, $\sigma$, and central velocities, $v_0$, which we refer to as the Cyg~OB2 (`c') and outlier (`o') populations. We introduce a free parameter, $f_o$, to represent the fraction of sources that are members of the outlier population and assume that the fraction of outliers is the same in both dimensions. While we don't know that the outliers are distributed according to a Gaussian (in fact there is very little reason to suppose that they are), the power of forward-modelling a Gaussian mixture model such as this comes not from making an accurate model of the outliers, but simply from acknowledging and modelling them \citep{hogg10}. To efficiently determine the best fitting parameters we use a Markov Chain Monte Carlo (MCMC) ensemble sampler \citep{fore13} to sample the posterior distribution function and maximise the likelihood function, assuming flat priors on all free parameters (the use of other priors does not change the results).

The best fitting parameters are $\sigma_\alpha(c) = 1.89^{+0.07}_{-0.06}$ and $\sigma_\delta(c) = 1.32^{+0.05}_{-0.04}$~mas/yr for the Cyg~OB2 component, and $\sigma_\alpha(o) = 36.9^{+4.7}_{-3.8}$ and $\sigma_\delta(o) = 32.4^{+4.1}_{-3.4}$~mas/yr for the outlier component, while the fraction of outliers is $f_o = 0.05 \pm 0.01$. We confirm that this more complex model has provided an improvement on the single component Gaussian model (which has fewer parameters) using the Bayesian Information Criterion \citep{schw78}, applying a penalty to the likelihood of the more complex two-component Gaussian model. Despite this penalty we find that the two-component Gaussian model provides a significantly improved fit compared to the single-component model.

The velocity dispersions (for the Cyg~OB2 component) obtained using this method are higher than those calculated using the uncertainty-corrected IQR method for $\sigma_\alpha$ and in good agreement with those for $\sigma_\delta$. This implies that while the PM distribution in the Declination dimension may be described as a Gaussian (once the outliers are accounted for), the distribution in the RA dimension is not Gaussian. The core of the $\mu_\alpha$ distribution is slightly under-fit by both velocity dispersion estimates, supporting the idea that a simple Gaussian does not fully represent the observed distribution. A more complex model could be used, though as we show in Section~\ref{s-vectormap} the underlying motions are highly substructured, meaning that such a model would have to be quite complex to accurately describe the kinematics of the system. Our goal here is to estimate the overall dynamical state of the system and in this sense these fits should be considered as reasonable approximations to the underlying velocity dispersion, albeit the best available approximation given the standard techniques for quantifying the velocity dispersion used here. We will use the forward-modelled velocity dispersions for the remainder of this work since they have included the most reliable treatment of the kinematic outliers and the PM uncertainties, and which we consider to be the most reliable and accurate estimate of the velocity dispersion.

\subsection{The 3-dimensional velocity dispersion}
\label{s-veldisp-3d}

Based on the distance to Cyg~OB2 of 1.33~kpc the velocity dispersions of the Cyg~OB2 component in our 2-component Gaussian mixture model are equivalent to $\sigma_\alpha(c) = 13.0^{+0.8}_{-0.7}$ and $\sigma_\delta(c) = 9.1^{+0.5}_{-0.5}$~km~s$^{-1}$ (confidence intervals include the uncertainty in the distance to Cyg~OB2). These values are similar to, but slightly larger than, the RV dispersion of the OB stars, $\sigma_{RV} = 8.03 \pm 0.26$~km~s$^{-1}$ \citep[using the mid-point RVs, $v_{mid} = 0.5 (v_{max} - v_{min})$, that are less susceptible to under-sampling than the average RVs,][]{kimi07,kimi08b}. Since the PM velocity dispersion of the OB stars is the same as that of the entire sample (Section~\ref{s-vdisp_mass}) we believe it acceptable to use the RV dispersion of the OB stars as representative of the overall RV dispersion.

While the velocity dispersions in the three dimensions are in approximate agreement with each other they are not identical, even within the confidence intervals. This implies that the velocity ellipsoid of Cyg~OB2 is not spherical (i.e. it is non-isotropic) and is actually triaxial with $\sigma_\alpha > \sigma_\delta > \sigma_{RV}$. The deviation from complete isotropy is not large and may be due to a dominant isotropic component within Cyg~OB2 combined with a minor component that is very strong along one axis. The direction of maximum velocity dispersion in the plane of the sky is at a position angle (PA) of 87.5$^\circ$. The fact that Cyg~OB2 does not have an isotropic velocity dispersion \citep[as is common for relaxed star clusters,][]{port10} supports previous suggestions that the association is not dynamically evolved \citep[e.g.,][]{wrig14b}.

Combining the velocity dispersions in each dimension we calculate the full 3 dimensional velocity dispersion as

\begin{equation}
\sigma_{3D}^2 = \langle v^2_{3D} \rangle = \sigma_\alpha^2 + \sigma^2_\delta + \sigma^2_{RV}
\end{equation}

\noindent
which gives $\sigma_{3D} = 17.8 \pm 0.6$~km~s$^{-1}$.

\subsection{Velocity dispersions as a function of stellar mass}
\label{s-vdisp_mass}

Gravitational encounters within a stellar system will drive the velocities of stars towards a thermal velocity distribution, whereby stars of different mass have the same energy, a state known as energy equipartition \citep{spit87}\footnote{Note that not all self-gravitating systems can attain complete energy equipartition \citep{spit69,tren13}.}. The development of equipartition within a cluster is also thought to generate mass segregation, the common observation that the most massive stars within a cluster or association are predominantly found in the densest parts of those regions \citep{hill98,goul04}. It is currently unknown as to whether mass segregation is due to some facet of the star formation process \citep[e.g.,][]{bonn98}, whether it can be rapidly induced through dynamical interactions \citep[e.g.,][]{alli09}, or whether the two processes can develop independently \citep[e.g.,][]{park16b,sper16}.

\begin{table}
\begin{center}
\caption{Velocity dispersion as a function of stellar mass in bins of 100 stars and calculated by forward modelling the IQR velocity dispersion of the total PM vector length ($\sqrt{\mu_\alpha^2 + \mu_\delta^2}$).}
\label{by_mass}
\begin{tabular}{cccc}
\hline
Median mass	& 68\% mass range		& \multicolumn{1}{c}{N$_\star$}		& $\langle \sigma^2 \rangle$	\\
($M_\odot$)	& ($M_\odot$)			&							& (mas/yr)$^2$				\\
\hline
10.7			& 8.15 -- 21.5		& 100			& $2.22_{-0.61}^{+0.88}$ \\
2.55			& 2.00 -- 5.90		& 100			& $1.28_{-0.40}^{+0.57}$ \\
1.80			& 1.80 -- 1.90		& 100			& $0.98_{-0.36}^{+0.48}$ \\
1.60			& 1.58 -- 1.70		& 100			& $2.04_{-0.67}^{+0.88}$ \\
1.30			& 1.30 -- 1.40		& 100			& $1.85_{-0.57}^{+0.77}$ \\
1.20			& 1.10 -- 1.20		& 100			& $2.28_{-0.74}^{+1.03}$ \\
1.00			& 0.89 -- 1.10		& 100			& $1.96_{-0.64}^{+0.90}$ \\
0.65			& 0.58 -- 0.76		& 100			& $2.04_{-0.79}^{+1.09}$ \\
0.45			& 0.37 -- 0.50		& 73				& $2.59_{-1.27}^{+1.95}$ \\
\hline
\end{tabular} 
\end{center}
\end{table}

To search for evidence of energy equipartition we studied the variation of the velocity dispersion as a function of stellar mass, dividing our sample into bins of 100 stars (73 stars for the final bin). For each bin we calculated the dispersion on the total vector lengths ($\sqrt{\mu_\alpha^2 + \mu_\delta^2}$), using the IQR to be resistant to outliers in these small samples, and forward modelling the distribution using a MCMC ensemble sampler to account for the impact of measurement uncertainties. It is particularly important to take into account these uncertainties since the estimated PM uncertainties are dependent on magnitude and will therefore increase as the stellar mass decreases. If the velocity dispersion is not properly corrected for uncertainties a false signal of energy equipartition may be observed when none is present.

\begin{figure}
\begin{center}
\includegraphics[height=240pt, angle=270]{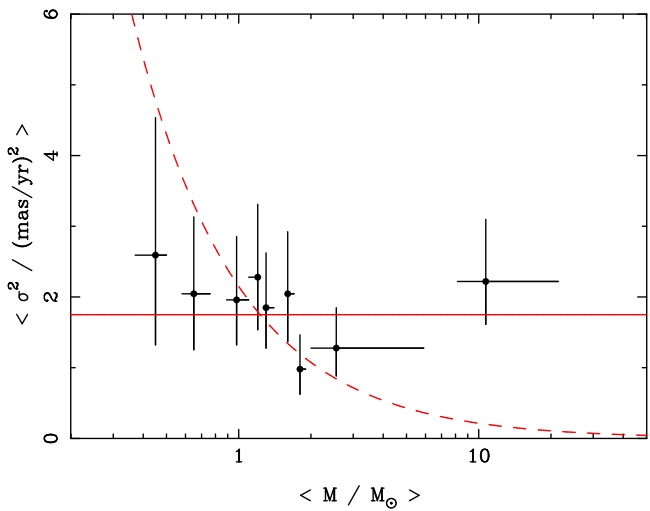}
\caption{PM velocity dispersion ($\langle \sigma^2 \rangle$) as a function of stellar mass in bins of 100 stars (see Table~\ref{by_mass}). The error bars show the 1$\sigma$ uncertainty in the velocity dispersions and the spread in mass for the stars in each bin. The solid line shows the best-fitting relationship expected if there were no energy equipartition ($\sigma^2 = \mathrm{constant}$), and the dashed line shows the best-fitting relationship expected if there were full energy equipartition ($\sigma^2 \propto M$).}
\label{energy_equipartition}
\end{center}
\end{figure}

The results are shown in Table~\ref{by_mass} and Figure~\ref{energy_equipartition}. The majority of data points are consistent with there being no energy equipartition, a result that does not change if the number of stars in each bin is changed. If the bin with the most massive stars is ignored there is a weak trend of decreasing velocity dispersion with increasing mass, but the results are also fully consistent with their being no energy equipartition. The most massive stars ($>$10~M$_\odot$) have a larger velocity dispersion than the intermediate-mass stars ($2-10$~M$_\odot$), and are more consistent with the low- and solar-mass stars ($0.4 - 2$~M$_\odot$). These results shouldn't be influenced by the presence of high-velocity kinematic outliers (due to our use of the IQR velocity dispersion) and cannot be caused by the broadening of the velocity dispersion by close binaries.

To determine the level of energy equipartition best represented by our data we adapted our two-component Gaussian mixture model to include a mass dependence on the velocity dispersion, which can be written as

\begin{equation}
\sigma_\alpha (m) \propto \sigma_\alpha(0) \, m^{-\eta}
\end{equation}

\begin{equation}
\sigma_\delta (m) \propto \sigma_\delta(0) \, m^{-\eta}
\end{equation}

\noindent where $\sigma_\alpha(0)$ and $\sigma_\delta(0)$ are the velocity dispersions in the two dimensions for $m = 1 M_\odot$, and $\eta$ is the degree of energy equipartition, where $\eta = 0$ is no energy equipartition and $\eta = 0.5$ is full energy equipartition. This introduces only one free parameter, $\eta$, compared to the model in Section~\ref{s-model}, because we assume the energy equipartition is isotropic and we do not apply energy equipartition to the outlier population. Using the ensemble MCMC sampler used above to sample the posterior distribution function gives a best fit of $\eta = 0.01 \pm 0.02$, consistent with no energy equipartition.

The lack of energy equipartition in Cyg~OB2 suggests the association has not undergone sufficient dynamical evolution for equipartition to occur. If energy equipartition is the main cause of mass segregation, the failure to detect the former implies little to none of the latter in Cyg~OB2, as is observed from the positions of stars of different masses within the association \citep{wrig14b}.

\subsection{What are the kinematic outliers?}
\label{s-outliers}

\begin{figure*}
\begin{center}
\includegraphics[height=500pt, angle=270]{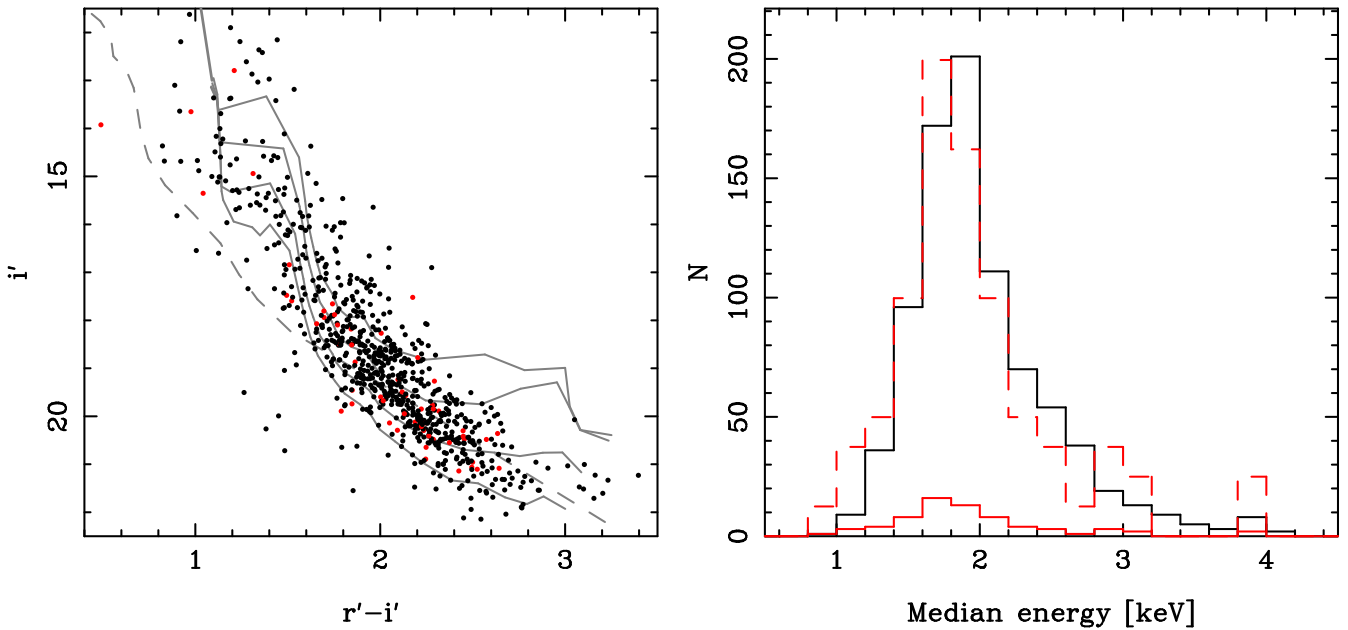}
\caption{The properties of kinematic outliers in our sample illustrating that there is no significant difference between the properties of the outliers and the entire sample. {\it Left:} An $i'$ vs $r'-i'$ colour-magnitude diagram showing 771 of the 873 sources in our sample (dots), including 59 out of 70 kinematic outliers (red dots). For reference the grey lines show pre-main sequence isochrones \citep{sies00} at ages of 1, 2, 5, and 10~Myr, at a distance of 1.33~kpc and an extinction of $A_V = 5.4$~mag \citep{wrig15a}, with IPHAS magnitudes calculated using the colours of \citet{drew05} and bolometric corrections from \citet{keny95}. The grey dashed line shows a foreground main-sequence at a distance of 500~pc \citep[a reasonable, representative distance for a mature X-ray emitting stellar population, see e.g.,][]{wrig10b} and an extinction of $A_V = 1.5$~mag \citep{sale14}. {\it Right:} The black histogram shows the distribution of median X-ray photon energies for the 848 X-ray sources in our sample, while the red solid histogram shows the distribution for the 70 kinematic outliers. The red dashed line shows the distribution of the kinematic outliers normalised to that of the entire sample distribution, showing a slight tendency towards lower median energies for the outliers.}
\label{outliers}
\end{center}
\end{figure*}

A small fraction of stars in our sample have relative PMs significantly larger than those of the bulk of the population. These were modelled in Section~\ref{s-model} as a separate outlier population and we estimated they represent $\sim$5\% of the entire sample. An important question is whether they stars are members of Cyg~OB2 that have been accelerated to large relative velocities by dynamical interactions, whether these are other young stars in the Cygnus region that have passed along the sightline towards Cyg~OB2 (but are otherwise not associated with the star formation events that formed the association), or whether they are contaminating, X-ray emitting field stars, most likely in the foreground. \citet{wrig10a} used IPHAS photometry to remove the majority of foreground contaminants down to $r^\prime = 20$~mag, though foreground sources fainter than this may remain in the sample.

To study the properties of the kinematic outliers relative to the rest of the sample we define as an outlier any star whose velocity, along either axis, deviates from the median by more than $3\sigma$, i.e. those that fulfil the criteria

\begin{equation}
|v_i-v_0| > 3 \sqrt{\sigma^2 + e_i^2}
\label{eqn-outliers}
\end{equation}

\noindent where $v_0$ and $\sigma$ are the central velocity and dispersion in that dimension, and $v_i$ and $e_i$ are the PM and uncertainty in the same dimension. By this definition we identify 70 kinematic outliers from our sample of 873 stars, or 8\% of the sample.

Figure~\ref{outliers} shows a colour magnitude diagram (CMD) for all stars with PMs, with the outliers indicated in red and appearing to broadly follow the same distribution as the entire sample. Figure~\ref{outliers} also shows representative pre-main-sequence isochrones and a foreground main sequence, illustrating that while the foreground and Cyg~OB2 populations would be marginally separated at the bright end of the CMD, they will overlap at the faint end. It is therefore difficult to determine whether the outliers are foreground contaminants based on their position in the CMD alone.

The median X-ray photon energy is a robust measure of the X-ray spectral shape that can provide information on the absorbing column and plasma temperature of the emitting material. Young stars, by virtue of being more X-ray luminous, have higher plasma temperatures and harder intrinsic X-ray spectra than contaminating field stars of the same mass \citep{tell05}. Furthermore, Cyg~OB2 members, being more distant, will lie behind a larger absorbing column of neutral hydrogen than the foreground contaminants, and since low energy X-ray photons are more readily absorbed than high energy X-ray photons, this will further harden the X-ray spectra of Cyg~OB2 members relative to foreground contaminants. Thus, Cyg~OB2 members will have larger median photon energies than foreground field star contaminants. \citet{getm11} simulated the X-ray spectral properties of foreground contaminants and found that 96\% had median energies less than 1.1~keV, while most young stars had median energies of 1--2.5~keV.

Figure~\ref{outliers} shows that the median X-ray photon energies for the kinematic outliers are very similar to the rest of the sample, with most sources having values of 1--3~keV, though there is a slight tendency for the outlier population to have lower median photon energies. This suggests the majority of the kinematic outliers are consistent with being X-ray absorbed young stars with possibly a small fraction of foreground contaminants. The young stars may include objects that have been ejected from Cyg~OB2, such as by the disruption of binary systems \citep[e.g.,][]{park14b}, or they may include stars that have dispersed from other nearby star formation sites \citep[see e.g.,][]{reip08}.

\section{Proper motion vector map}
\label{s-vectormap}

\begin{figure*}
\begin{center}
\includegraphics[width=500pt]{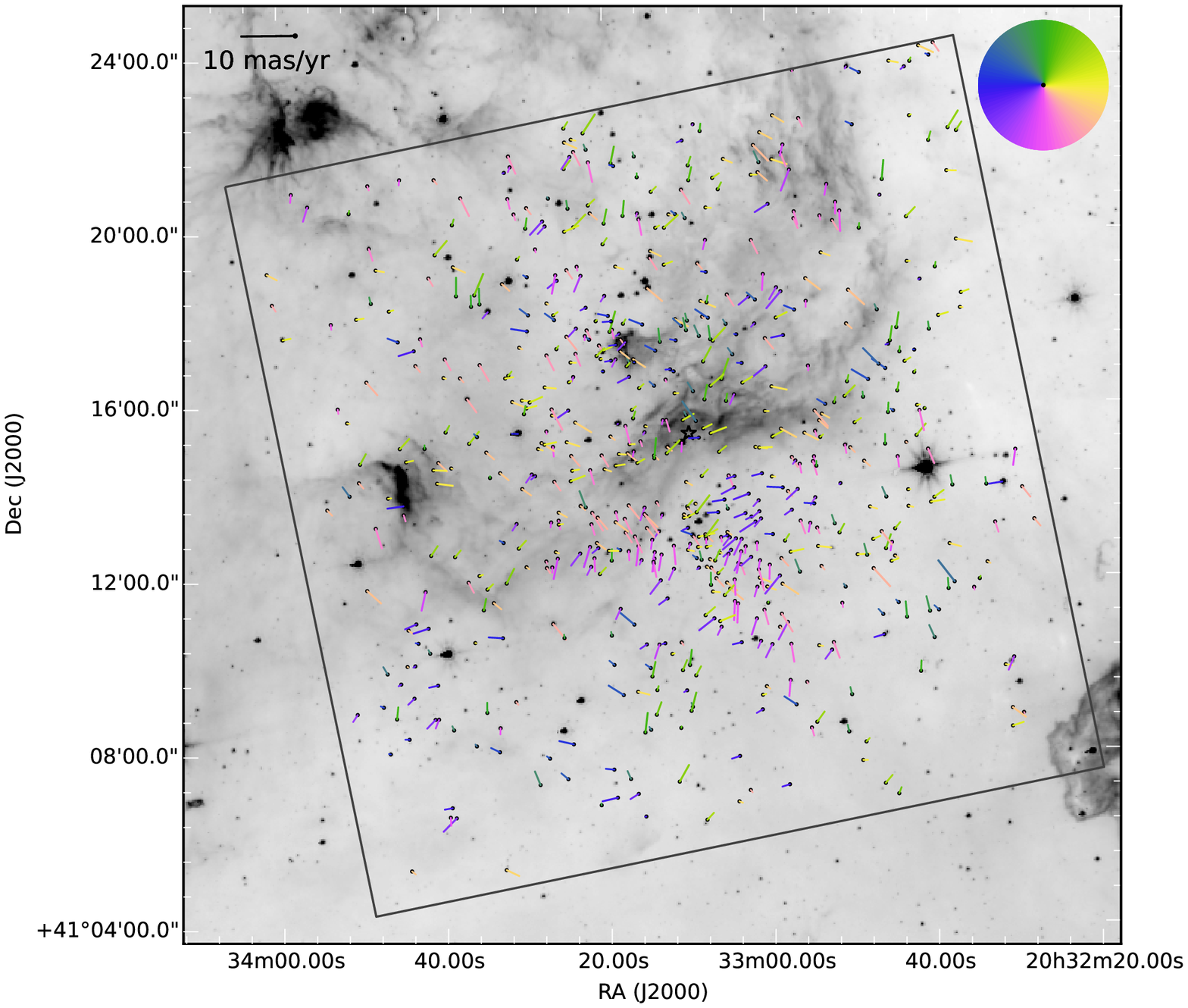}
\caption{PM vector map for 798 X-ray and spectroscopically selected stars towards Cygnus~OB2 including 16 O-type stars, 34 B-type stars, and 748 X-ray selected stars. The 75 most extreme kinematic outliers, as noted in the text, have been removed. The dots show the current position of the stars, while the vectors shown the PMs, colour-coded based on their direction of motion to highlight the kinematic substructure. The grey box shows the border of the X-ray observations used to identify members of Cyg~OB2 and an empty black star symbol marks the centre of mass of the association as determined in Section~\ref{s-expconrot}. A representative 10~mas/yr vector is shown in the top left corner and a colour wheel showing the relationship between colour and PA is shown in the top right corner. The background is a {\it Spitzer} 8~$\mu$m image \citep{hora07}.}
\label{vectormap}
\end{center}
\end{figure*}

Here we present the PM vector map (Figure~\ref{vectormap}) and discuss the bulk stellar motions. For this analysis we have removed from our sample 75  kinematic outliers, 70 fulfilling Eqn.~\ref{eqn-outliers}) and 5 with uncertainties $>5$~mas/yr in either dimension. We also note briefly that the plane of the sky at the Galactic longitude of Cyg~OB2 ($l \sim 80^\circ$) is almost perpendicular to the direction of Galactic rotation and therefore the measured PMs carry only a negligible component due to it.

The stellar motions shown in Figure~\ref{vectormap} appear to be quite random, particularly on the largest scales, with no clear sign of expansion or contraction. On smaller scales there is some evidence for coherent motions, with stars in the same area of the association moving in approximately the same direction. This is often referred to as {\it kinematic substructure}, particularly in RV studies, though it is much more apparent here than in previous works \citep[e.g.,][]{jeff14,tobi15}. In this Section we discuss these features in more detail, beginning in Section~\ref{s-expconrot} with a quantification of the level of expansion, contraction and rotation in the association and then in Section~\ref{s-kinsubstructure} the evidence for kinematic substructure is appraised.

\subsection{Expansion, contraction and rotation}
\label{s-expconrot}

\begin{figure*}
\begin{center}
\includegraphics[height=500pt, angle=270]{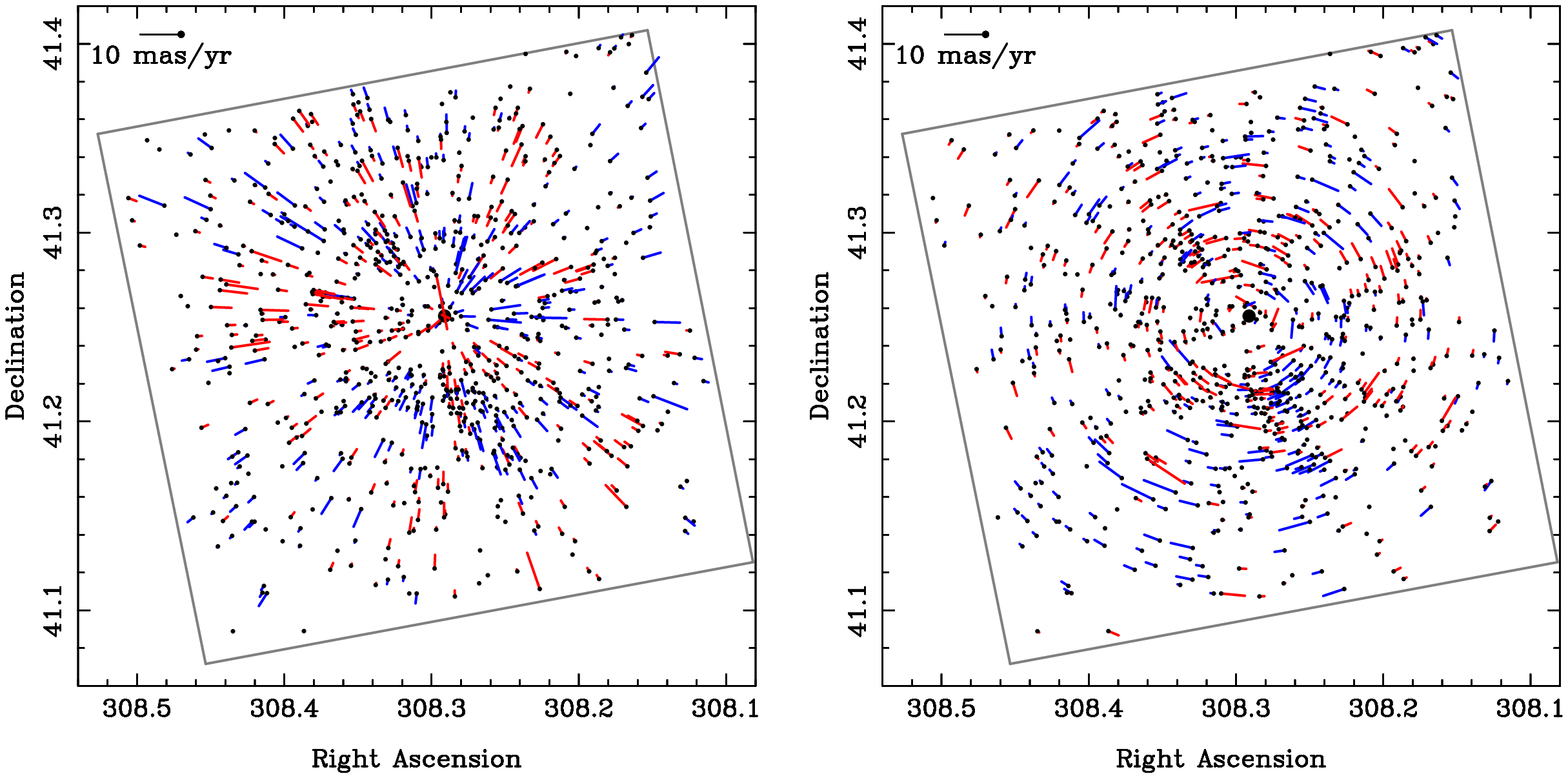}
\caption{Split-component PM vector maps for the 798 members of Cyg~OB2 (excluding kinematic outliers) as shown in Figure~\ref{vectormap}. In both panels the dots show the current positions of the stars, the vectors show the PMs, and the large black dot shows the nominal centre of the association. The left-hand panel shows the radial component of the PM vectors, colour-coded blue if the stars are moving outwards from the centre, and red if they are moving inwards. The right-hand panel shows the transverse component of the vectors, colour-coded blue if the stars are moving in a clockwise direction and red if they are moving in an anti-clockwise direction.}
\label{expansion_vectors}
\end{center}
\end{figure*}

To examine these components of motion, we separate the PM vectors into radial and azimuthal components (as shown in Figure~\ref{expansion_vectors}) and use stellar masses from \citet{wrig10a} and \citet{wrig15a} to calculate the amount of kinetic energy in each component (uncertainties estimated using a Monte Carlo simulation).

To do this we must estimate the centre of the association, which is not a simple task for highly substructured OB associations such as Cyg~OB2. We explored various possibilities for the centre of the association, seeking to identify the position that maximised the amount of kinetic energy in either rotational or azimuthal directions, but found that these quantities exhibited only a weak dependence on the choice of association centre. We therefore chose the centre of Cyg~OB2 to be the centre of mass of the sample used here \citep[considering only stars in the mass ranges identified by][for which the sample is believed to be spatially unbiased]{wrig14b}, which is 20:33:10, +41:15:21. This is approximately half-way between the two estimates in the literature, the trapezium of bright O stars that make up Cyg~OB2 \#8 \citep[e.g.,][]{hans03,vink08} and the centre of the association found from an infrared star counts study by \citet{knod00}. It is also very close to the centre of mass of the OB star sample from \citet{wrig15a}.

We find that the kinetic energy is divided between the radial and azimuthal components in the ratio $60_{-7}^{+3}$:$40_{-3}^{+7}$, suggesting a slight ($\sim$1.5$\sigma$ confidence) preference for kinetic energy in the radial direction over the azimuthal direction. This result holds approximately true regardless of the centre of the association used \citep[the ratio varies from 63:37 when using the centre of mass of the OB stars, to 58:42 when using the centre determined by][]{knod00}, or whether the entire PM sample is used or only those stars in the mass ranges considered complete (for which the ratio is 62:38).

In the radial direction there is an almost even split in both the number of expanding and contracting stars ($51_{-1}^{+2}$:$49_{-2}^{+1}$) and the kinetic energy ($50_{-7}^{+9}$:$50_{-9}^{+7}$) in both expansion and contraction (i.e. away from or towards the centre of the association), a result that shows very little variation when different centres or subsets of the sample are considered. When using the centre of mass of the OB stars the ratio of expanding to contracting energies changes to 43:57, the largest variation seen, and none of the centres result in more than half of the kinetic energy being in expansion.

In the azimuthal direction there is a preference for motion in the direction of decreasing PA with $66_{-7}^{+5}$\% of the azimuthal kinetic energy in that direction and $34_{-5}^{+7}$\% in the direction of increasing PA (this result is independent of the centre used). A similar split is seen in the distribution of angular momentum with $61_{-4}^{+2}$\% in the direction of decreasing PA and $39_{-2}^{+4}$\% in the direction of increasing PA. Since the number of stars moving in each azimuthal direction and their mass distributions are very similar, these difference must be entirely due to the stars moving faster in the direction of decreasing PA. If this were a gravitationally bound system this would be evidence of rotation, but because Cyg~OB2 is not bound (and may never have been bound, see Section~\ref{s-discussion}) it is more accurate to refer to this as non-zero angular momentum, and is most likely a remnant of the angular momentum of the primordial GMC \citep[e.g.,][]{roso03}.

\subsection{Kinematic substructure}
\label{s-kinsubstructure}

While the overall kinematic structure appears to be relatively random, with no evidence for cohesive expanding or contracting motions, the small-scale kinematics suggests some substructure. The PM vectors in Figure~\ref{vectormap} have been colour-coded based on their PA to highlight this. We use the term \emph{kinematic substructure} to describe the observed tendency for stars in the same area of the association to have similar PMs to their neighbours, both in direction and in magnitude. This is evident on a wide range of scales, from that of only a few stars, up to groups of 10--20 stars or more, and appears to exist across the OB association. Kinematic substructure has been observed or hinted at in a small number of past kinematic studies of star forming regions and star clusters, but it is considerably more apparent in these PM observations than in past RV studies \citep[e.g.,][]{fure08,jeff14,tobi15}.

To determine whether this apparent kinematic substructure is real or whether it is a chance fluctuation we need to quantify its significance. We do this using spatial correlation tests, which are designed to search for correlations in a signal among nearby locations in space. Global indexes of spatial correlation, such as Moran's I \citep{mora50} and Geary's C \citep{gear54}, express the overall degree of similarity between spatially close regions with respect to a numeric variable. Both tests involve computing a degree of similarity, $\rho_\mu$, between every possible pair of points, $i$ and $j$, with respect to the numerical variable of interest, $\mu$, which in our case would be the PM along one axis. All the values of $\rho_\mu$ are then summed up, weighted by the degree of proximity, $w_{ij}$, between points $i$ and $j$, and then divided by a constant of proportionality. The resulting index reveals whether the data are consistent with a random distribution, or whether it displays significant evidence of positive (nearby regions will tend to exhibit similar values of $\mu$) or negative (nearby regions exhibit dissimilar values of $\mu$) spatial correlation. The two indexes differ slightly in that Moran's I statistic is a global measure of spatial correlation, while Geary's C statistic is a more local measure of correlation.

Moran's I statistic is given by

\begin{equation}
I = \frac{n}{\sum\limits_{i=1}^n \sum\limits_{j=1}^n w_{ij} } \, \frac{ \sum\limits_{i=1}^n \sum\limits_{j=1}^n w_{ij} (\mu_i - \bar{\mu}) (\mu_j - \bar{\mu})}{ \sum\limits_{i=1}^n (\mu_i - \bar{\mu})^2 }
\end{equation}

\noindent where the degree of similarity in this case is $\rho_\mu = (\mu_i - \bar{\mu}) (\mu_j - \bar{\mu})$, $\bar{\mu}$ is the mean of $\mu$, and $n$ is the number of data points. We use the standard weighting of $w_{ij} = 1 / d_{ij}$, where $d_{ij}$ is the distance between $i$ and $j$. Under the null hypothesis of no spatial correlation the expected value is $I_0 = -1 / (n-1)$. Values of $I > I_0$ indicate positive spatial correlation, while $I < I_0$ indicates negative correlation. The variance of $I$ can be calculated using either the normal approximation \citep{mora50} or by randomisation experiments, though for large sample sizes ($n > 25$) they are very similar and the normal approximation is sufficient \citep{upto85}.

Using the PM in each dimension as the variable of interest ($\mu$), we calculate values of $I_\alpha = 0.024 \pm 0.0026$ and $I_\delta = 0.031 \pm 0.0026$. Both values deviate significantly from the expectation value under the null hypothesis of $I_0 = -0.00125$ with significances of 9.7 and 12.5$\sigma$ respectively, implying that there is significant positive spatial correlation in our sample, i.e., the PMs are spatially correlated with stars close to each other having more similar values than for a random distribution.

We also calculated the degree of spatial correlation using Geary's C statistic, which is given by

\begin{equation}
C = \frac{n-1}{2 \sum\limits_{i=1}^n \sum\limits_{j=1}^n w_{ij}} \, \frac{ \sum\limits_{i=1}^n \sum\limits_{j=1}^n w_{ij} (\mu_i - \mu_j)^2 }{ \sum\limits_{i=1}^n (\mu_i - \bar{\mu})^2 }
\end{equation}

\noindent where the degree of similarity here is $\rho_\mu = (\mu_i - \mu_j)^2$ and the same weighting is used as with Moran's I statistic. The expectation value for no spatial correlation is $C_0 = 1$, with lower values implying positive spatial correlation and higher values meaning negative spatial correlation. As with Moran's I statistic the variance was calculated using the normal approximation \citep{gear54}.

We calculate values of $C_\alpha = 0.964 \pm 0.014$ and $C_\delta = 0.951 \pm 0.014$, both of which imply positive spatial correlation. The significance of these results are calculated as 2.6 and 3.5$\sigma$, both statistically significant, but lower than the results from Moran's I statistic. This is most likely due to Geary's C statistics being a more local measure of correlation, indicating there are probably large areas where the local correlation is low (i.e., don't exhibit kinematic substructure) and small areas that are highly correlated (have strong kinematic substructure). Despite this both measures of spatial correlation indicate that the PMs of stars in Cyg~OB2 exhibit statistically significant positive spatial correlation in the form of kinematic substructure.

It is important to confirm that the substructures observed are real and not due to artefacts, the most likely cause of which would be correlations between the uncertainties in $\mu_\alpha$ and $\mu_\delta$ introduced by either the data reduction process \citep[see e.g.,][for an example of this for Hipparcos data]{perr98} or by atmospheric turbulence on large scales. Unfortunately it is difficult to directly test this because there isn't a suitable reference catalogue free from such uncertainties. However, if correlated uncertainties did exist within our data and were responsible for inducing false kinematic features into our PMs then such features should also be evident in the kinematics of non-member sources.

To investigate this we studied the kinematics of non-members in the same area of the sky as our sample of Cyg~OB2 members (a total of $\sim$15,000 sources with PM uncertainties $<$5~mas/yr), but could find no patterns or substructures in their distribution. To quantify this we used the two spatial correlation tests used earlier to search for evidence of kinematic substructure within our sample of non-members. We created 10,000 bootstrapped samples of 798 non-members by randomly selecting non-member stars within 1$^\prime$ of each member star. A radius of 1$^\prime$ was chosen to allow a sufficiently large sample of non-members to sample from whilst also ensuring that the spatial distribution of our bootstrapped samples was similar to our sample of Cyg~OB2 members. For each sample we calculated Geary's C statistic and Moran's I statistic for both $\mu_\alpha$ and $\mu_\delta$.

\begin{figure}
\begin{center}
\includegraphics[height=240pt, angle=270]{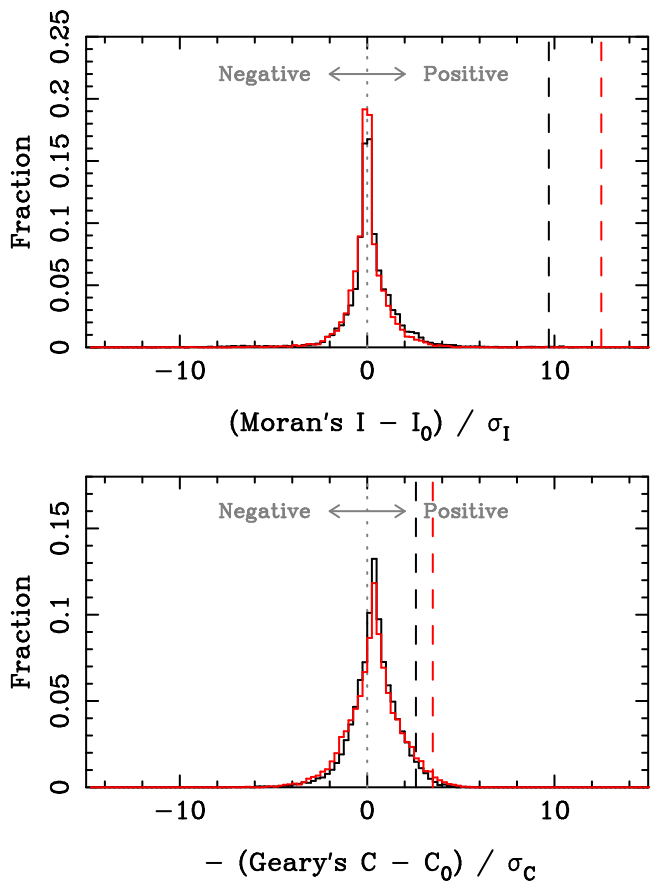}
\caption{The distribution of the significances of Geary's C and Moran's I test statistics for non-members of Cyg~OB2, calculated as the test statistic minus the expectation value and then divided by the standard deviation for each sample (necessary because the expectation value and standard deviation vary with the properties of each sample). For Geary's C statistic we multiplied the values by $-1$ so that both distributions show positive values for positive spatial correlation and negative values for negative spatial correlation. The black histogram shows the distribution for $\mu_\alpha$ and the red histogram shows that for $\mu_\delta$. The vertical dashed lines show the values measured for our sample of Cyg~OB2 members.}
\label{spatial_test_statistics}
\end{center}
\end{figure}

Figure~\ref{spatial_test_statistics} shows the distribution of the significances of the two spatial correlation tests. Since the expectation value and standard deviation of each test statistic vary with each sample we calculated the significance of each measurement and plotted this distribution. All the distributions are narrow and centred on zero (or very close to zero for Geary's C statistic), implying little to no spatial correlation in the kinematics. The spatial correlation significances measured for Cyg~OB2 members are also shown in Figure~\ref{spatial_test_statistics}. For Geary's C statistic the two measures are within the tail of the distribution, while for Moran's I statistic they are well outside of the distribution. This suggests that the spatially correlated PMs observed in Cyg~OB2 are not a product of the observations or the data reduction process and are therefore real.

\section{Discussion}
\label{s-discussion}

Here we discuss the implications of our results on our current understanding of Cyg~OB2 and its dynamical state. Our results can be briefly summarised as follows.

\begin{itemize}
\item The PM velocity dispersions are $\sigma_\alpha(c) = 13.0^{+0.8}_{-0.7}$ and $\sigma_\delta(c) = 9.1^{+0.5}_{-0.5}$~km~s$^{-1}$, which are non-isotropic. Combined with the RV dispersion this gives a 3-dimensional velocity dispersion of $\sigma_{3D} = 17.8 \pm 0.6$~km~s$^{-1}$.
\item There is no evidence for energy equipartition in the stellar kinematics, implying that the association is not dynamically evolved, a picture supported by the lack of mass segregation in the association \citep{wrig14b}.
\item The PMs do not display a global expansion pattern and the kinetic energy in expanding (outwards) motion is roughly the same as that in contracting (inwards) motion.
\item There is roughly the same amount of kinetic energy in the azimuthal and radial directions, with a slight preference in the former for motion in the direction of decreasing PA.
\item The PMs exhibit considerable kinematic substructure that is evident from pairs of stars with very similar kinematics all the way up to larger groups of many tens of stars moving together. This echoes the physical substructure already known in the association \citep{wrig14b}.
\end{itemize}

We now consider the implications of these results for the dynamical history of Cyg~OB2, including theories for the origin of OB associations.

\subsection{The dynamical state of Cyg OB2}

To determine the virial state of Cyg~OB2 we use the virial equation, which in its 3-dimensional form is given by

\begin{equation}
\sigma_{3D}^2 = \frac{ G M_{vir} }{ 2 r_{vir} } ,
\end{equation}

\noindent where $\sigma_{3D}$ is the 3-dimensional velocity dispersion (equal to $17.8 \pm 0.6$~km~s$^{-1}$), $G$ is the gravitational constant, $M_{vir}$ is the virial mass and $r_{vir}$ is the virial radius. Following convention we substitute the parameter $\eta = 6 r_{vir} / r_{eff}$, where $r_{eff}$ is the effective (or half-light) radius \citep[e.g.,][]{port10}. Rearranging this gives $M_{vir}$ as

\begin{equation}
M_{vir} = \eta \frac{ \sigma_{3D}^2 r_{eff}}{3 G} .
\end{equation}

\noindent The parameters $\eta$ and $r_{eff}$ are determined by fitting an \citet{elso87} surface brightness profile to the stellar distribution, which has the form

\begin{equation}
\Sigma (r) = \Sigma_0 \left( 1 + \frac{r^2}{a^2} \right)^{-\gamma/2} ,
\end{equation}

\noindent where $\Sigma$ is the stellar surface density, $r$ is the projected radial distance from the association centre, and $a$ and $\gamma$ are parameters to be fit. Using the sample of O-B0 stars from \citet{wrig15a} and the centre of mass of the association calculated in Section~\ref{s-expconrot} we find parameters of $\gamma = 5.8 \pm 0.5$ and $a = 19.4 \pm 1.9$\arcm. These parameters correspond to $\eta = 9.7 \pm 0.8$ and $r_{eff} = 10.1 \pm 0.9$\arcm\ \citep{port10}, the latter of which equates to 3.9~pc at a distance of 1.33~kpc.

This gives a virial mass of $(9.3 \pm 0.8) \times 10^5$~M$_\odot$. The total stellar mass has been estimated to be between $1.65^{+0.38}_{-0.28} \times 10^4$~M$_\odot$ \citep{wrig15a} and $(4 - 10) \times 10^4$~M$_\odot$ \citep{knod00}, with most estimates placing it around $(2-4) \times 10^4$~M$_\odot$ \citep[e.g.,][]{drew08,wrig10a}. The virial mass is therefore over an order of magnitude larger than the stellar mass, implying that Cyg~OB2 is gravitationally unbound. This result is unchanged if we adopt our (smaller) IQR velocity dispersion, which gives $\sigma_{3D} = 13.9 \pm 0.4$~km~s$^{-1}$ and therefore $M_{vir} = (5.7 \pm 0.6) \times 10^5$~M$_\odot$.

This calculation does not take into account any gas embedded within the association. However this will be minimal because Cyg~OB2 is not embedded within a molecular cloud but in a cavity between the two major parts of the Cygnus~X GMC \citep[e.g.,][]{schn06}. A first-order estimate of the mass of intra-cluster gas is obtained from

\begin{equation}
M_{gas} = \mu_{H_2} \, m_H \int N_H dA \simeq \mu_{H_2} \, m_H N_H A
\end{equation}

\noindent where $\mu_{H_2}$ is the atomic mass of molecular hydrogen, $m_H$ is the mass of a hydrogen atom, $N_H$ is the hydrogen column density through the association, and $A$ is its projected area. Assuming a single value of $N_H$ estimated from the extinction through the association \citep[$\Delta A_V \sim 3$~mags, see Figure~3 of][]{wrig15a}, and the conversion of \citet{bohl78}\footnote{Since the reddening law towards and through Cyg~OB2 appears to be normal \citep{hans03,wrig15a} there is no reason to think that a typical gas to dust ratio might be incorrect.} we derive $N_H = 2.8 \times 10^{21}$~cm$^{-2}$. We estimate the projected area of Cyg~OB2 to be a circle with a radius twice the effective radius, equating to $A = 0.053$~deg$^2$. This gives a total intra-cluster gas mass of 1300~M$_\odot$, which will not significantly contribute to the virial equation.

These calculations confirm that Cyg~OB2 is gravitationally unbound, at least globally, as the majority of OB associations are believed to be \citep{amba51,blaa64}. The association must therefore be in the process of expanding and dispersing into the galactic field, which it will do in a very short period of time given its highly super-virial state. Based on the 1-dimensional velocity dispersion of $\sim$10~km~s$^{-1}$, and given that 1~km~s$^{-1}$ $\sim$ 1~pc~Myr$^{-1}$, this means that the association will expand by approximately 10~pc in radius per Myr. Within 3-4 Myr the association will have expanded to be over 100~pc across, roughly equivalent in size to the Scorpius-Centaurus OB association \citep{prei08}.

\subsection{Is Cyg OB2 an expanded star cluster?}

The classical view of OB associations is that they are the expanded remnants of disrupted star clusters \citep[e.g.,][]{brow99,lada03}. The physical processes suggested for the disruption are either residual gas expulsion \citep[in which feedback disperses the gas left over from star formation, which was previously holding the cluster in virial equilibrium, e.g.,][]{hill80,lada84,good06} or tidal heating from nearby molecular clouds \citep{spit58,elme10,krui11b}. If this were the case it would imply that Cyg~OB2 was denser and more compact in the past and has since expanded to become a low-density OB association.

We would therefore expect the PMs to exhibit either radial expanding motions (for the explosive expansion predicted by residual gas expulsion) or expanding motions along a specific axis \citep[see Figure~3 of][for an illustration]{krui11}. While the full space motions for stars in Cyg~OB2 are not yet available (given the absence of RVs for the lower mass stars), their PMs do not exhibit any sort of correlated expansion pattern. There is also no preference for kinetic energy in the radial direction, as expected if it were expanding from its apparent centre. This evidence therefore rules out Cyg~OB2 having been a dense star cluster in the past.

This conclusion is supported by other evidence that Cyg~OB2 is not dynamically evolved, as would be expected if it had previously been a dense and compact star cluster. This includes a lack of mass segregation \citep{wrig14b}, energy equipartition (Section~\ref{s-vdisp_mass}), or an isotropic velocity dispersion\footnote{It is possible tidal heating could generate a non-isotropic velocity dispersion on a global scale, due to the preferential orientation on which it acts, though the lack of a signature in the PM vector diagram argues against this.} (Section~\ref{s-veldisp}), all of which are indicators of a dynamically evolved system \citep{port10}. The considerable physical \citep{wrig14b} and kinematic (Section~\ref{s-kinsubstructure}) substructure also suggests the association is not dynamically mixed, otherwise this substructure would have been erased \citep{park14}.

The lack of a clear expansion pattern in Cyg~OB2 would appear to be in conflict with the measured velocity dispersion that suggests the association is gravitationally unbound. A possible explanation for this is that the association has only recently become gravitationally unbound and therefore has not dispersed sufficiently to develop a clear radial expansion pattern. This appears unlikely given that the association has already dispersed its primordial molecular cloud \citep[see e.g., Figure~1 of][]{wrig15a}, unless the molecular cloud was dispersed very rapidly, perhaps by a particularly powerful supernova. Another possibility is that we are seeing Cyg~OB2 at this current time as a chance overdensity of many substructures that have overlapped along the line of sight, in which case the expansion of the OB association as a whole is a rather meaningless concept given that it would never have been a single structure.

\subsection{Properties of the kinematic substructures}
\label{s-properties}

Determining the precise properties and virial states of all the individual substructures is beyond the scope of this paper, partly because of the difficulty identifying individual groups and assigning stars to them. We reserve a full analysis of the size, structure and virial state of these subgroups for a future paper, but briefly estimate their properties here.

To calculate the typical masses of these structures we first estimate the number of stars observed in each group as between 10--100 members. We know that our sample is approximately complete for $M > 1 M_\odot$ \citep{wrig10a} and that $\sim$70\% of the stars in our sample have masses $>$1~M$_\odot$, meaning that these groups contain approximately 7--70 stars in this mass range. In a typical and fully sampled initial mass function, stars with masses $>$1~M$_\odot$ represent about 10\% of the total number of stars and the mean stellar mass is 0.6~M$_\odot$ \citep{masc13}. Therefore 10--100 stars in our sample is approximately equivalent to 40--400~M$_\odot$, a reasonable estimate for the typical mass of these groups.

The larger groups (identified roughly by eye) appear to have velocity dispersions that are consistent, within the uncertainties, with being in virial equilibrium (based on an extrapolation of the observed stars to a fully sampled initial mass function). This is supported by the fact that these groups are still moving together and are not noticeably expanding. If these groups were gravitationally unbound it is reasonable to expect that they would have dispersed in the 3--5~Myr since they formed. The high overall velocity dispersion of Cyg~OB2 is probably a superposition of all the subgroups, each of which might be bound or close to virial equilibrium, but have mean velocities slightly offset from one another such that the overall dispersion is a wide Gaussian.

None of the groups correspond to the two open clusters identified by \citet{bica03} in the centre of Cyg~OB2. The brightest stars in these clusters don't have PMs in our sample (they are saturated), while the other stars do not have any coherent kinematic structure in our PMs. It is therefore difficult to verify the nature of these clusters.

\subsection{What impact has residual gas expulsion had on the dynamics of Cyg~OB2?}

The lack of a radial expansion in the global kinematic structure suggests that residual gas expulsion and other cluster disruption mechanisms were not responsible for globally unbinding Cyg~OB2 (despite the fact that the association has expelled the majority of its residual gas). Furthermore, if the kinematic substructures are in (or close to) virial equilibrium (Section~\ref{s-properties}), then such cluster disruption mechanisms have also had very little impact on the virial state of these structures. It is possible that some of the stars that appear isolated in phase-space (i.e., that are not part of any substructure) may be remnants of an expanded cluster, but there are no obvious trends in their kinematics to verify this.

If residual gas expulsion really has had little impact on the dynamics of Cyg~OB2 it would be in agreement with a number of recent theoretical studies. \citet{krui12b} found that in hydrodynamic simulations of star formation stars accrete the majority of gas in their local vicinity and thus groups of stars carve out regions of the molecular cloud free from gas, reducing any dynamical impact arising from its expulsion. \citet{moec12} and \citet{dale12} both argued that small clusters of stars have short-enough dynamical timescales that they can settle into virialised and stellar-dominated configurations before feedback begins, potentially allowing them to survive gas expulsion. \citet{dale12} also showed that the densest parts of molecular clouds can survive considerable ionization, limiting the extent to which residual gas is expelled. Future, higher-precision kinematic measurements (e.g., from Gaia) will allow these ideas to be tested in more detail.

\subsection{The possible origin and evolution of the kinematic groups}

Recent infrared and sub-mm observations have shown that stars form in a highly substructured distribution \citep[e.g.,][]{gute08} with both spatial and kinematic subclustering \citep{test00} that is thought to arise from the filamentary structure of the primordial molecular cloud \citep{andr14,rath15}. If this is a universal aspect of the star formation process then the kinematic groups we observe in Cyg~OB2 may be the remnants of this substructure. What these substructures might be able to tell us about Cyg~OB2 depends on how much these groups have evolved since the stars within them formed.

Numerical simulations show that substructure in star forming regions is rapidly erased by dynamical interactions between groups \citep[e.g.,][]{scal02,park14}, and that if the stars are sufficiently dynamically cool (subvirial) these interactions lead to the formation of dense star clusters by hierarchical mergers \citep{alli10,fuji13}. These simulations also suggest that if the inter-group dynamics are hot (supervirial) then the groups will separate, preserving the substructure \citep{good04}. In this framework the hierarchical merging of groups in a star forming region proceeds until they reach the physical scale at which the inter-group dynamics transition from subvirial to supervirial. If this scale is larger than the entire region then a star cluster forms, while if it is smaller then a substructured OB association forms. If this picture is correct it would explain our observations of Cyg~OB2 that show a globally unbound association composed of kinematic substructures that are at (or very close to) virial equilibrium. One could hypothesise that the structures we observe formed from mergers between smaller substructures, but that the supervirial inter-group dynamics prevented further mergers.

These simulations suggest that substructures grow by dynamical interactions, but cannot separate into smaller structures \citep{park14}. This would imply that the structures observed in Cyg~OB2 place constraints on the largest dynamically mixed structures that have ever existed within the association. However, there may be dynamical processes that are capable of rapidly eroding groups of stars. Processes such as three-body encounters \citep{pove67} or supernovae explosions \citep{blaa61} are thought to be responsible for ejecting individual stars, but it might be possible that the ejection of particularly massive stars could disrupt the local stellar dynamics sufficiently to eject multiple stars. Supernova explosions could also rapidly redistribute the gas in the vicinity of the association (and therefore the gas potential), allowing processes such as tidal heating \citep{elme10} to strip off large numbers of stars from otherwise bound groups. The kinematic signatures of such a complicated and disruptive event might be very difficult to identify.

There are also a number of stars in Cyg~OB2 that do not appear to be part of any moving group, either because they are in a sparse area of the association, or because their kinematics are very different from those of nearby stars. These stars may have been born in relative isolation, they may have been stripped off from other moving groups, or they may have originated in a group or cluster that has been completely dispersed. Their PMs do not suggest they all originated in a single cluster that was disrupted (their motions appear relatively random and do not exhibit radial dispersion), but they might have originated in multiple clusters.

It may be possible with future observations to assign the majority of stars in Cyg~OB2 into one kinematic group or another, in which case we can speculate as to what this could tell us about star formation in Cyg~OB2. If we assume that cluster evaporation and stripping is not a rapid process \citep{port10} and that these groups have evolved predominantly through mergers \citep{scal02}, then the kinematic groups we observe are equivalent to the largest dynamically mixed structures that have existed in Cyg~OB2. Some studies have claimed that the mass of a cluster correlates with the mass of its most massive star \citep{weid04,weid06}. Cyg~OB2 contains two stars with masses of $\sim$100~M$_\odot$ \citep{wrig15a}, which according to this framework must have formed in a cluster of mass $\sim$10,000~M$_\odot$. This is over an order of magnitude larger than the largest dynamically mixed structures currently observed in the association. Even if the groups we observed have experienced significant disruptive mass loss they would be unlikely to have been as massive as $\sim$10,000~M$_\odot$ in the past. This argues against there being a relationship between cluster mass and the mass of the most massive star within it \citep[see also][]{park07,cerv13}.

\section{Conclusions}

We have presented a catalogue of 873 high-precision PMs for X-ray and spectroscopically selected members of the massive OB association Cyg~OB2. The PMs were calculated from several thousand images spanning $\sim$15~years and from various instruments on multiple telescopes. The PMs are accurate to $\sim$0.5~mas/yr for sources brighter than $i \sim 18$~mag and to better than 1~mas/yr for sources at $i \sim 22$~mag.

The velocity distribution of Cyg~OB2 members is broadly Gaussian with low-amplitude and broad wings. We model this using a 2-dimensional, 2-component Gaussian mixture model representing the Cyg~OB2 and kinematic outlier components. For the Cyg~OB2 component we calculate velocity dispersions of $1.89^{+0.07}_{-0.06}$ and $1.32^{+0.05}_{-0.04}$~mas/yr in the two PM dimensions using this method, in good agreement with that derived from simpler outlier-resistant analytical methods. The kinematic outliers, which represent approximately 5\% of our sample, appear to be predominantly young stars at the approximate distance of Cyg~OB2 (probably including ejected stars and unrelated stars from nearby star forming regions) and very few appear to be foreground stars.

At the distance of Cyg~OB2 the velocity dispersion is equivalent to $\sigma_\alpha(c) = 13.0^{+0.8}_{-0.7}$ and $\sigma_\delta(c) = 9.1^{+0.5}_{-0.5}$~km~s$^{-1}$, which combined with the RV dispersion gives a 3-dimensional velocity dispersion of $\sigma_{3D} = 17.8 \pm 0.6$~km~s$^{-1}$. This implies a virial mass an order of magnitude larger than the observed stellar mass, implying that Cyg~OB2 is gravitationally unbound.

The PMs exhibit significant kinematic structure, echoing the observed physical substructure in the association \citep{wrig14b}. The kinematic substructure implies the association is dynamically unevolved, a view supported by a lack of energy equipartition and non-isotropic velocity dispersions. The kinematic substructures appear to be close to virial equilibrium, and have typical sizes of $\sim$40--400~M$_\odot$.

The PMs show no evidence for a global expansion pattern, as would be expected if the association was an expanded star cluster that had been disrupted by mechanisms such as residual gas expulsion or tidal heating. Furthermore, since the substructures appear to be in (or close to) virial equilibrium, this suggests that disruption mechanisms such as residual gas expulsion have had very little impact on the dynamical state of the association or its substructures.

These results all suggest that Cyg~OB2 was not born as a single dense star cluster and instead was born with considerable physical and kinematic substructure, much of which has survived to the present day. The classical view of OB associations as the expanded remnants of star clusters disrupted by residual gas expulsion does not appear to be valid for Cyg~OB2.

These results could be tested using RVs, which would allow the kinematic substructure and lack of energy equipartition to be verified in a third dimension. Higher-precision PMs (e.g., from the Gaia satellite) could be used to determine the virial state of the substructures we have identified, and therefore to constrain their past evolution. The evolution of complex substructured regions such as this could also be tested in more detail by making comparisons with the results of N-body simulations \citep[e.g.,][]{pros09,park12b}, particularly those that facilitate quantitative comparisons between observations and simulations through the use of well-defined spatial and kinematic diagnostics \citep{alli09,park16}.

\section{Acknowledgments}

We thank the anonymous referee for a swift and insightful report that improve this paper. NJW acknowledges a Royal Astronomical Society Research Fellowship and an STFC Ernest Rutherford Fellowship. HB is funded by the Ram\'on y Cajal fellowship program number RYC-2009-04497 and by Spanish Grant AYA2012-38897-C02-01.  DB was funded by Spanish grant AYA2012-38897-C02-01. The authors would like to thank Nate Bastian, Simon Goodwin, Diederik Kruijssen, Estelle Moraux, and Richard Parker for comments on this paper.

This paper makes use of data obtained from the Canada-France-Hawaii Telescope (operated by the National Research Council of Canada, the Centre National de la Recherche Scientifique of France, and the University of Hawaii), the Isaac Newton Telescope (operated by the Isaac Newton Group in the Spanish Observatorio del Roque de los Muchachos), the Two Micron All Sky Survey (a joint project of the University of Massachusetts and the Infrared Processing and Analysis Center/California Institute of Technology), the UKIRT Infrared Deep Sky Survey, the Sloan Digital Sky Survey (funding provided by the Alfred P. Sloan Foundation), and the Kiso Observatory (operated by the University of Tokyo). 
Data for this paper was obtained using the Isaac Newton Group Archive (maintained as part of the CASU Astronomical Data Centre at the Institute of Astronomy, Cambridge), the Canadian Astronomy Data Centre (operated by the National Research Council of Canada with the support of the Canadian Space Agency), the NOAO Science Archive (operated by the Association of Universities for Research in Astronomy under agreement with the National Science Foundation), and SMOKA (operated by the Astronomy Data Center, National Astronomical Observatory of Japan). 
This research has made use of the Vizier and SIMBAD databases (operated at CDS, Strasbourg, France).

\bibliographystyle{mn2e}
\bibliography{/Users/nwright/Documents/Work/tex_papers/bibliography.bib}

\begin{thebibliography}{126}
\expandafter\ifx\csname natexlab\endcsname\relax\def\natexlab#1{#1}\fi

\bibitem[{{Adams} {et~al}\mbox{.}(2006){Adams}, {Proszkow}, {Fatuzzo}, \&
  {Myers}}]{adam06}
{Adams} F.~C., {Proszkow} E.~M., {Fatuzzo} M., {Myers} P.~C., 2006, \apj, 641,
  504

\bibitem[{{Allison} {et~al}\mbox{.}(2010){Allison}, {Goodwin}, {Parker},
  {Portegies Zwart}, \& {de Grijs}}]{alli10}
{Allison} R.~J., {Goodwin} S.~P., {Parker} R.~J., {Portegies Zwart} S.~F., {de
  Grijs} R., 2010, \mnras, 407, 1098

\bibitem[{{Allison} {et~al}\mbox{.}(2009){Allison}, {Goodwin}, {Parker},
  {Portegies Zwart}, {de Grijs}, \& {Kouwenhoven}}]{alli09}
{Allison} R.~J., {Goodwin} S.~P., {Parker} R.~J., {Portegies Zwart} S.~F., {de
  Grijs} R., {Kouwenhoven} M.~B.~N., 2009, \mnras, 395, 1449

\bibitem[{{Ambarzumjan}(1951)}]{amba51}
{Ambarzumjan} W.~A., 1951, in Abhandlungen Aus Der Sowjetischen Astronomie,
  Folge I, {Singer} O., ed., p.~33

\bibitem[{{Andr{\'e}} {et~al}\mbox{.}(2014){Andr{\'e}}, {Di Francesco},
  {Ward-Thompson}, {Inutsuka}, {Pudritz}, \& {Pineda}}]{andr14}
{Andr{\'e}} P., {Di Francesco} J., {Ward-Thompson} D., {Inutsuka} S.-I.,
  {Pudritz} R.~E., {Pineda} J.~E., 2014, Protostars and Planets VI, 27

\bibitem[{{Bailer-Jones} {et~al}\mbox{.}(2000){Bailer-Jones}, {Bizenberger}, \&
  {Storz}}]{bail00}
{Bailer-Jones} C.~A., {Bizenberger} P., {Storz} C., 2000, in Society of
  Photo-Optical Instrumentation Engineers (SPIE) Conference Series, Vol. 4008,
  Optical and IR Telescope Instrumentation and Detectors, {Iye} M., {Moorwood}
  A.~F., eds., pp. 1305--1316

\bibitem[{{Bastian} {et~al}\mbox{.}(2007){Bastian}, {Ercolano}, {Gieles},
  {Rosolowsky}, {Scheepmaker}, {Gutermuth}, \& {Efremov}}]{bast07}
{Bastian} N., {Ercolano} B., {Gieles} M., {Rosolowsky} E., {Scheepmaker} R.~A.,
  {Gutermuth} R., {Efremov} Y., 2007, \mnras, 379, 1302

\bibitem[{{Baumgardt} \& {Kroupa}(2007)}]{baum07}
{Baumgardt} H., {Kroupa} P., 2007, \mnras, 380, 1589

\bibitem[{{Bertin}(2006)}]{bert06}
{Bertin} E., 2006, in Astronomical Society of the Pacific Conference Series,
  Vol. 351, Astronomical Data Analysis Software and Systems XV, {Gabriel} C.,
  {Arviset} C., {Ponz} D., {Enrique} S., eds., p. 112

\bibitem[{{Bertin}(2011)}]{bert11}
{Bertin} E., 2011, in Astronomical Society of the Pacific Conference Series,
  Vol. 442, Astronomical Data Analysis Software and Systems XX, {Evans} I.~N.,
  {Accomazzi} A., {Mink} D.~J., {Rots} A.~H., eds., p. 435

\bibitem[{{Bertin} \& {Arnouts}(1996)}]{bert96}
{Bertin} E., {Arnouts} S., 1996, \aaps, 117, 393

\bibitem[{{Bica} {et~al}\mbox{.}(2003){Bica}, {Bonatto}, \& {Dutra}}]{bica03}
{Bica} E., {Bonatto} C., {Dutra} C.~M., 2003, \aap, 405, 991

\bibitem[{{Blaauw}(1961)}]{blaa61}
{Blaauw} A., 1961, {Bulletin of the Astronomical Institutes of the
  Netherlands}, 15, 265

\bibitem[{{Blaauw}(1964)}]{blaa64}
{Blaauw} A., 1964, \araa, 2, 213

\bibitem[{{Bohlin} {et~al}\mbox{.}(1978){Bohlin}, {Savage}, \&
  {Drake}}]{bohl78}
{Bohlin} R.~C., {Savage} B.~D., {Drake} J.~F., 1978, \apj, 224, 132

\bibitem[{{Bonnell} {et~al}\mbox{.}(2001){Bonnell}, {Bate}, {Clarke}, \&
  {Pringle}}]{bonn01}
{Bonnell} I.~A., {Bate} M.~R., {Clarke} C.~J., {Pringle} J.~E., 2001, \mnras,
  323, 785

\bibitem[{{Bonnell} \& {Davies}(1998)}]{bonn98}
{Bonnell} I.~A., {Davies} M.~B., 1998, \mnras, 295, 691

\bibitem[{{Bonnell} {et~al}\mbox{.}(2011){Bonnell}, {Smith}, {Clark}, \&
  {Bate}}]{bonn11}
{Bonnell} I.~A., {Smith} R.~J., {Clark} P.~C., {Bate} M.~R., 2011, \mnras, 410,
  2339

\bibitem[{{Boulade} {et~al}\mbox{.}(2003){Boulade}, {Charlot}, {Abbon}, {Aune},
  {Borgeaud}, {Carton}, {Carty}, {Da Costa}, {Deschamps}, {Desforge},
  {Eppell{\'e}}, {Gallais}, {Gosset}, {Granelli}, {Gros}, {de Kat}, {Loiseau},
  {Ritou}, {Rouss{\'e}}, {Starzynski}, {Vignal}, \& {Vigroux}}]{boul03}
{Boulade} O. {et~al.}, 2003, in Society of Photo-Optical Instrumentation
  Engineers (SPIE) Conference Series, Vol. 4841, Instrument Design and
  Performance for Optical/Infrared Ground-based Telescopes, {Iye} M.,
  {Moorwood} A.~F.~M., eds., pp. 72--81

\bibitem[{{Bouy} {et~al}\mbox{.}(2013){Bouy}, {Bertin}, {Moraux}, {Cuillandre},
  {Bouvier}, {Barrado}, {Solano}, \& {Bayo}}]{bouy13}
{Bouy} H., {Bertin} E., {Moraux} E., {Cuillandre} J.-C., {Bouvier} J.,
  {Barrado} D., {Solano} E., {Bayo} A., 2013, \aap, 554, A101

\bibitem[{{Bressert} {et~al}\mbox{.}(2010){Bressert}, {Bastian}, {Gutermuth},
  {Megeath}, {Allen}, {Evans}, {Rebull}, {Hatchell}, {Johnstone}, {Bourke},
  {Cieza}, {Harvey}, {Merin}, {Ray}, \& {Tothill}}]{bres10}
{Bressert} E. {et~al.}, 2010, \mnras, 409, L54

\bibitem[{{Brown} {et~al}\mbox{.}(1999){Brown}, {Blaauw}, {Hoogerwerf}, {de
  Bruijne}, \& {de Zeeuw}}]{brow99}
{Brown} A.~G.~A., {Blaauw} A., {Hoogerwerf} R., {de Bruijne} J.~H.~J., {de
  Zeeuw} P.~T., 1999, in NATO Advanced Science Institutes (ASI) Series C, Vol.
  540, NATO Advanced Science Institutes (ASI) Series C, {Lada} C.~J., {Kylafis}
  N.~D., eds., p. 411

\bibitem[{{Brown} {et~al}\mbox{.}(1997){Brown}, {Dekker}, \& {de
  Zeeuw}}]{brow97}
{Brown} A.~G.~A., {Dekker} G., {de Zeeuw} P.~T., 1997, \mnras, 285, 479

\bibitem[{{Carpenter}(2000)}]{carp00}
{Carpenter} J.~M., 2000, \aj, 120, 3139

\bibitem[{{Casali} {et~al}\mbox{.}(2007){Casali}, {Adamson}, {Alves de
  Oliveira}, {Almaini}, {Burch}, {Chuter}, {Elliot}, {Folger}, {Foucaud},
  {Hambly}, {Hastie}, {Henry}, {Hirst}, {Irwin}, {Ives}, {Lawrence}, {Laidlaw},
  {Lee}, {Lewis}, {Lunney}, {McLay}, {Montgomery}, {Pickup}, {Read}, {Rees},
  {Robson}, {Sekiguchi}, {Vick}, {Warren}, \& {Woodward}}]{casa07}
{Casali} M. {et~al.}, 2007, \aap, 467, 777

\bibitem[{{Cervi{\~n}o} {et~al}\mbox{.}(2013){Cervi{\~n}o},
  {Rom{\'a}n-Z{\'u}{\~n}iga}, {Luridiana}, {Bayo}, {S{\'a}nchez}, \&
  {P{\'e}rez}}]{cerv13}
{Cervi{\~n}o} M., {Rom{\'a}n-Z{\'u}{\~n}iga} C., {Luridiana} V., {Bayo} A.,
  {S{\'a}nchez} N., {P{\'e}rez} E., 2013, \aap, 553, A31

\bibitem[{{Comer{\'o}n} {et~al}\mbox{.}(2002){Comer{\'o}n}, {Pasquali},
  {Rodighiero}, {Stanishev}, {De Filippis}, {L{\'o}pez Mart{\'{\i}}},
  {G{\'a}lvez Ortiz}, {Stankov}, \& {Gredel}}]{come02}
{Comer{\'o}n} F. {et~al.}, 2002, \aap, 389, 874

\bibitem[{{Dale} {et~al}\mbox{.}(2012){Dale}, {Ercolano}, \&
  {Bonnell}}]{dale12}
{Dale} J.~E., {Ercolano} B., {Bonnell} I.~A., 2012, \mnras, 424, 377

\bibitem[{{de Bruijne}(2012)}]{debr12}
{de Bruijne} J.~H.~J., 2012, \apss, 341, 31

\bibitem[{{Drew} {et~al}\mbox{.}(2005){Drew}, {Greimel}, {Irwin},
  {Aungwerojwit}, {Barlow}, {Corradi}, {Drake}, {G{\"a}nsicke}, {Groot},
  {Hales}, {Hopewell}, {Irwin}, \& {Knigge}}]{drew05}
{Drew} J.~E. {et~al.}, 2005, \mnras, 362, 753

\bibitem[{{Drew} {et~al}\mbox{.}(2008){Drew}, {Greimel}, {Irwin}, \&
  {Sale}}]{drew08}
{Drew} J.~E., {Greimel} R., {Irwin} M.~J., {Sale} S.~E., 2008, \mnras, 386,
  1761

\bibitem[{{Elmegreen}(2002)}]{elme02}
{Elmegreen} B.~G., 2002, \apj, 564, 773

\bibitem[{{Elmegreen}(2008)}]{elme08}
{Elmegreen} B.~G., 2008, \apj, 672, 1006

\bibitem[{{Elmegreen} \& {Elmegreen}(2001)}]{elme01}
{Elmegreen} B.~G., {Elmegreen} D.~M., 2001, \aj, 121, 1507

\bibitem[{{Elmegreen} \& {Hunter}(2010)}]{elme10}
{Elmegreen} B.~G., {Hunter} D.~A., 2010, \apj, 712, 604

\bibitem[{{Elson} {et~al}\mbox{.}(1987){Elson}, {Fall}, \& {Freeman}}]{elso87}
{Elson} R.~A.~W., {Fall} S.~M., {Freeman} K.~C., 1987, \apj, 323, 54

\bibitem[{{F{\H u}r{\'e}sz} {et~al}\mbox{.}(2008){F{\H u}r{\'e}sz}, {Hartmann},
  {Megeath}, {Szentgyorgyi}, \& {Hamden}}]{fure08}
{F{\H u}r{\'e}sz} G., {Hartmann} L.~W., {Megeath} S.~T., {Szentgyorgyi} A.~H.,
  {Hamden} E.~T., 2008, \apj, 676, 1109

\bibitem[{{Foreman-Mackey} {et~al}\mbox{.}(2013){Foreman-Mackey}, {Hogg},
  {Lang}, \& {Goodman}}]{fore13}
{Foreman-Mackey} D., {Hogg} D.~W., {Lang} D., {Goodman} J., 2013, \pasp, 125,
  306

\bibitem[{{Foster} {et~al}\mbox{.}(2015){Foster}, {Cottaar}, {Covey}, {Arce},
  {Meyer}, {Nidever}, {Stassun}, {Tan}, {Chojnowski}, {da Rio}, {Flaherty},
  {Rebull}, {Frinchaboy}, {Majewski}, {Skrutskie}, {Wilson}, \&
  {Zasowski}}]{fost15}
{Foster} J.~B. {et~al.}, 2015, \apj, 799, 136

\bibitem[{{Fujii} \& {Portegies Zwart}(2013)}]{fuji13}
{Fujii} M.~S., {Portegies Zwart} S., 2013, \mnras, 430, 1018

\bibitem[{{Geary}(1954)}]{gear54}
{Geary} R.~C., 1954, The Incorporated Statistician, 5, 115

\bibitem[{{Getman} {et~al}\mbox{.}(2011){Getman}, {Broos}, {Feigelson},
  {Townsley}, {Povich}, {Garmire}, {Montmerle}, {Yonekura}, \&
  {Fukui}}]{getm11}
{Getman} K.~V. {et~al.}, 2011, \apjs, 194, 3

\bibitem[{{Goodwin} \& {Bastian}(2006)}]{good06}
{Goodwin} S.~P., {Bastian} N., 2006, \mnras, 373, 752

\bibitem[{{Goodwin} \& {Whitworth}(2004)}]{good04}
{Goodwin} S.~P., {Whitworth} A.~P., 2004, \aap, 413, 929

\bibitem[{{Gouliermis} {et~al}\mbox{.}(2004){Gouliermis}, {Keller}, {Kontizas},
  {Kontizas}, \& {Bellas-Velidis}}]{goul04}
{Gouliermis} D., {Keller} S.~C., {Kontizas} M., {Kontizas} E., {Bellas-Velidis}
  I., 2004, \aap, 416, 137

\bibitem[{{Guarcello} {et~al}\mbox{.}(2012){Guarcello}, {Wright}, {Drake},
  {Garc{\'{\i}}a-Alvarez}, {Drew}, {Aldcroft}, \& {Kashyap}}]{guar12}
{Guarcello} M.~G., {Wright} N.~J., {Drake} J.~J., {Garc{\'{\i}}a-Alvarez} D.,
  {Drew} J.~E., {Aldcroft} T., {Kashyap} V.~L., 2012, \apjs, 202, 19

\bibitem[{{Gutermuth} {et~al}\mbox{.}(2008){Gutermuth}, {Myers}, {Megeath},
  {Allen}, {Pipher}, {Muzerolle}, {Porras}, {Winston}, \& {Fazio}}]{gute08}
{Gutermuth} R.~A. {et~al.}, 2008, \apj, 674, 336

\bibitem[{{Hanson}(2003)}]{hans03}
{Hanson} M.~M., 2003, \apj, 597, 957

\bibitem[{{Hillenbrand} {et~al}\mbox{.}(1998){Hillenbrand}, {Strom}, {Calvet},
  {Merrill}, {Gatley}, {Makidon}, {Meyer}, \& {Skrutskie}}]{hill98}
{Hillenbrand} L.~A., {Strom} S.~E., {Calvet} N., {Merrill} K.~M., {Gatley} I.,
  {Makidon} R.~B., {Meyer} M.~R., {Skrutskie} M.~F., 1998, \aj, 116, 1816

\bibitem[{{Hills}(1980)}]{hill80}
{Hills} J.~G., 1980, \apj, 235, 986

\bibitem[{{Hogg} {et~al}\mbox{.}(2010){Hogg}, {Bovy}, \& {Lang}}]{hogg10}
{Hogg} D.~W., {Bovy} J., {Lang} D., 2010, ArXiv e-prints

\bibitem[{{Hora} {et~al}\mbox{.}(2007){Hora}, {Bontemps}, {Megeath},
  {Schneider}, {Motte}, {Carey}, {Simon}, {Keto}, {Smith}, {Allen},
  {Gutermuth}, {Fazio}, {Kraemer}, {Mizuno}, {Price}, \& {Adams}}]{hora07}
{Hora} J. {et~al.}, 2007, in {Massive Star Formation: Observations Confront
  Theory}

\bibitem[{{Ives}(1998)}]{ives98}
{Ives} D., 1998, IEEE Spectrum, 16, 20

\bibitem[{{Ivezic} {et~al}\mbox{.}(2008){Ivezic}, {Axelrod}, {Brandt}, {Burke},
  {Claver}, {Connolly}, {Cook}, {Gee}, {Gilmore}, {Jacoby}, {Jones}, {Kahn},
  {Kantor}, {Krabbendam}, {Lupton}, {Monet}, {Pinto}, {Saha}, {Schalk},
  {Schneider}, {Strauss}, {Stubbs}, {Sweeney}, {Szalay}, {Thaler}, {Tyson}, \&
  {LSST Collaboration}}]{ivez08}
{Ivezic} Z. {et~al.}, 2008, Serbian Astronomical Journal, 176, 1

\bibitem[{{Jeffries} {et~al}\mbox{.}(2014){Jeffries}, {Jackson}, {Cottaar},
  {Koposov}, {Lanzafame}, {Meyer}, {Prisinzano}, {Randich}, {Sacco},
  {Brugaletta}, {Caramazza}, {Damiani}, {Franciosini}, {Frasca}, {Gilmore},
  {Feltzing}, {Micela}, {Alfaro}, {Bensby}, {Pancino}, {Recio-Blanco}, {de
  Laverny}, {Lewis}, {Magrini}, {Morbidelli}, {Costado}, {Jofr{\'e}},
  {Klutsch}, {Lind}, \& {Maiorca}}]{jeff14}
{Jeffries} R.~D. {et~al.}, 2014, \aap, 563, A94

\bibitem[{{Kenyon} \& {Hartmann}(1995)}]{keny95}
{Kenyon} S.~J., {Hartmann} L., 1995, \apjs, 101, 117

\bibitem[{{Kiminki} {et~al}\mbox{.}(2007){Kiminki}, {Kobulnicky}, {Kinemuchi},
  {Irwin}, {Fryer}, {Berrington}, {Uzpen}, {Monson}, {Pierce}, \&
  {Woosley}}]{kimi07}
{Kiminki} D.~C. {et~al.}, 2007, \apj, 664, 1102

\bibitem[{{Kiminki} {et~al}\mbox{.}(2008){Kiminki}, {Kobulnicky}, {Kinemuchi},
  {Irwin}, {Fryer}, {Berrington}, {Uzpen}, {Monson}, {Pierce}, \&
  {Woosley}}]{kimi08b}
{Kiminki} D.~C. {et~al.}, 2008, \apj, 681, 735

\bibitem[{{Kiminki} {et~al}\mbox{.}(2015){Kiminki}, {Kobulnicky}, {Vargas
  {\'A}lvarez}, {Alexander}, \& {Lundquist}}]{kimi15}
{Kiminki} D.~C., {Kobulnicky} H.~A., {Vargas {\'A}lvarez} C.~A., {Alexander}
  M.~J., {Lundquist} M.~J., 2015, ArXiv e-prints

\bibitem[{{Kn{\"o}dlseder}(2000)}]{knod00}
{Kn{\"o}dlseder} J., 2000, \aap, 360, 539

\bibitem[{{Kroupa} {et~al}\mbox{.}(2001){Kroupa}, {Aarseth}, \&
  {Hurley}}]{krou01}
{Kroupa} P., {Aarseth} S., {Hurley} J., 2001, \mnras, 321, 699

\bibitem[{{Kruijssen}(2011)}]{krui11}
{Kruijssen} J.~M.~D., 2011, in Stellar Clusters and Associations: A RIA
  Workshop on Gaia, pp. 137--141

\bibitem[{{Kruijssen}(2012)}]{krui12}
{Kruijssen} J.~M.~D., 2012, \mnras, 426, 3008

\bibitem[{{Kruijssen} {et~al}\mbox{.}(2012){Kruijssen}, {Maschberger},
  {Moeckel}, {Clarke}, {Bastian}, \& {Bonnell}}]{krui12b}
{Kruijssen} J.~M.~D., {Maschberger} T., {Moeckel} N., {Clarke} C.~J., {Bastian}
  N., {Bonnell} I.~A., 2012, \mnras, 419, 841

\bibitem[{{Kruijssen} {et~al}\mbox{.}(2011){Kruijssen}, {Pelupessy}, {Lamers},
  {Portegies Zwart}, \& {Icke}}]{krui11b}
{Kruijssen} J.~M.~D., {Pelupessy} F.~I., {Lamers} H.~J.~G.~L.~M., {Portegies
  Zwart} S.~F., {Icke} V., 2011, \mnras, 414, 1339

\bibitem[{{Krumholz}(2014)}]{krum14}
{Krumholz} M.~R., 2014, Physics Reports, 539, 49

\bibitem[{{Lada} \& {Lada}(2003)}]{lada03}
{Lada} C.~J., {Lada} E.~A., 2003, \araa, 41, 57

\bibitem[{{Lada} {et~al}\mbox{.}(1984){Lada}, {Margulis}, \&
  {Dearborn}}]{lada84}
{Lada} C.~J., {Margulis} M., {Dearborn} D., 1984, \apj, 285, 141

\bibitem[{{Lamb} {et~al}\mbox{.}(2010){Lamb}, {Oey}, {Werk}, \&
  {Ingleby}}]{lamb10}
{Lamb} J.~B., {Oey} M.~S., {Werk} J.~K., {Ingleby} L.~D., 2010, \apj, 725, 1886

\bibitem[{{Lucas} {et~al}\mbox{.}(2008){Lucas}, {Hoare}, {Longmore},
  {Schr{\"o}der}, {Davis}, {Adamson}, {Bandyopadhyay}, {de Grijs}, {Smith}, \&
  {Gosling}}]{luca08}
{Lucas} P.~W. {et~al.}, 2008, \mnras, 391, 136

\bibitem[{{Maschberger}(2013)}]{masc13}
{Maschberger} T., 2013, \mnras, 429, 1725

\bibitem[{{Massey} \& {Thompson}(1991)}]{mass91}
{Massey} P., {Thompson} A.~B., 1991, \aj, 101, 1408

\bibitem[{{Moeckel} {et~al}\mbox{.}(2012){Moeckel}, {Holland}, {Clarke}, \&
  {Bonnell}}]{moec12}
{Moeckel} N., {Holland} C., {Clarke} C.~J., {Bonnell} I.~A., 2012, \mnras, 425,
  450

\bibitem[{{Moran}(1950)}]{mora50}
{Moran} P.~A.~P., 1950, Biometrika, 37, 17

\bibitem[{{Parker} \& {Goodwin}(2007)}]{park07}
{Parker} R.~J., {Goodwin} S.~P., 2007, \mnras, 380, 1271

\bibitem[{{Parker} {et~al}\mbox{.}(2016){Parker}, {Goodwin}, {Wright}, {Meyer},
  \& {Quanz}}]{park16b}
{Parker} R.~J., {Goodwin} S.~P., {Wright} N.~J., {Meyer} M.~R., {Quanz} S.~P.,
  2016, \mnras

\bibitem[{{Parker} \& {Meyer}(2012)}]{park12b}
{Parker} R.~J., {Meyer} M.~R., 2012, \mnras, 427, 637

\bibitem[{{Parker} \& {Meyer}(2014)}]{park14b}
{Parker} R.~J., {Meyer} M.~R., 2014, \mnras, 442, 3722

\bibitem[{{Parker} \& {Wright}(2016)}]{park16}
{Parker} R.~J., {Wright} N.~J., 2016, \mnras, {Submitted}

\bibitem[{{Parker} {et~al}\mbox{.}(2014){Parker}, {Wright}, {Goodwin}, \&
  {Meyer}}]{park14}
{Parker} R.~J., {Wright} N.~J., {Goodwin} S.~P., {Meyer} M.~R., 2014, \mnras,
  438, 620

\bibitem[{{Perryman} {et~al}\mbox{.}(1998){Perryman}, {Brown}, {Lebreton},
  {Gomez}, {Turon}, {Cayrel de Strobel}, {Mermilliod}, {Robichon},
  {Kovalevsky}, \& {Crifo}}]{perr98}
{Perryman} M.~A.~C. {et~al.}, 1998, \aap, 331, 81

\bibitem[{{Perryman} {et~al}\mbox{.}(2001){Perryman}, {de Boer}, {Gilmore},
  {H{\o}g}, {Lattanzi}, {Lindegren}, {Luri}, {Mignard}, {Pace}, \& {de
  Zeeuw}}]{perr01}
{Perryman} M.~A.~C. {et~al.}, 2001, \aap, 369, 339

\bibitem[{{Pfalzner}(2009)}]{pfal09}
{Pfalzner} S., 2009, \aap, 498, L37

\bibitem[{{Pflamm-Altenburg} \& {Kroupa}(2009)}]{pfla09}
{Pflamm-Altenburg} J., {Kroupa} P., 2009, \mnras, 397, 488

\bibitem[{{Portegies Zwart} {et~al}\mbox{.}(2010){Portegies Zwart}, {McMillan},
  \& {Gieles}}]{port10}
{Portegies Zwart} S.~F., {McMillan} S.~L.~W., {Gieles} M., 2010, \araa, 48, 431

\bibitem[{{Poveda} {et~al}\mbox{.}(1967){Poveda}, {Ruiz}, \& {Allen}}]{pove67}
{Poveda} A., {Ruiz} J., {Allen} C., 1967, Boletin de los Observatorios
  Tonantzintla y Tacubaya, 4, 86

\bibitem[{{Preibisch} \& {Feigelson}(2005)}]{prei05}
{Preibisch} T., {Feigelson} E.~D., 2005, \apjs, 160, 390

\bibitem[{{Preibisch} \& {Mamajek}(2008)}]{prei08}
{Preibisch} T., {Mamajek} E., 2008, {The Nearest OB Association:
  Scorpius-Centaurus (Sco OB2)}, {Handbook of Star Forming Regions}, p. 235

\bibitem[{{Proszkow} {et~al}\mbox{.}(2009){Proszkow}, {Adams}, {Hartmann}, \&
  {Tobin}}]{pros09}
{Proszkow} E.-M., {Adams} F.~C., {Hartmann} L.~W., {Tobin} J.~J., 2009, \apj,
  697, 1020

\bibitem[{{Puget} {et~al}\mbox{.}(2004){Puget}, {Stadler}, {Doyon}, {Gigan},
  {Thibault}, {Luppino}, {Barrick}, {Benedict}, {Forveille}, {Rambold},
  {Thomas}, {Vermeulen}, {Ward}, {Beuzit}, {Feautrier}, {Magnard}, {Mella},
  {Preis}, {Vallee}, {Wang}, {Lin}, {Hall}, \& {Hodapp}}]{puge04}
{Puget} P. {et~al.}, 2004, in Society of Photo-Optical Instrumentation
  Engineers (SPIE) Conference Series, Vol. 5492, Ground-based Instrumentation
  for Astronomy, {Moorwood} A.~F.~M., {Iye} M., eds., pp. 978--987

\bibitem[{{Rathborne} {et~al}\mbox{.}(2015){Rathborne}, {Longmore}, {Jackson},
  {Alves}, {Bally}, {Bastian}, {Contreras}, {Foster}, {Garay}, {Kruijssen},
  {Testi}, \& {Walsh}}]{rath15}
{Rathborne} J.~M. {et~al.}, 2015, \apj, 802, 125

\bibitem[{{Reipurth} \& {Schneider}(2008)}]{reip08}
{Reipurth} B., {Schneider} N., 2008, {Star Formation and Young Clusters in
  Cygnus}, {Handbook of Star Forming Regions}, p.~36

\bibitem[{{Robichon} {et~al}\mbox{.}(1999){Robichon}, {Arenou}, {Mermilliod},
  \& {Turon}}]{robi99}
{Robichon} N., {Arenou} F., {Mermilliod} J.-C., {Turon} C., 1999, \aap, 345,
  471

\bibitem[{{Rosolowsky} {et~al}\mbox{.}(2003){Rosolowsky}, {Engargiola},
  {Plambeck}, \& {Blitz}}]{roso03}
{Rosolowsky} E., {Engargiola} G., {Plambeck} R., {Blitz} L., 2003, \apj, 599,
  258

\bibitem[{{Rygl} {et~al}\mbox{.}(2012){Rygl}, {Brunthaler}, {Sanna}, {Menten},
  {Reid}, {van Langevelde}, {Honma}, {Torstensson}, \& {Fujisawa}}]{rygl12}
{Rygl} K.~L.~J. {et~al.}, 2012, \aap, 539, A79

\bibitem[{{Sale} {et~al}\mbox{.}(2014){Sale}, {Drew}, {Barentsen}, {Farnhill},
  {Raddi}, {Barlow}, {Eisl{\"o}ffel}, {Vink}, {Rodr{\'{\i}}guez-Gil}, \&
  {Wright}}]{sale14}
{Sale} S.~E. {et~al.}, 2014, \mnras, 443, 2907

\bibitem[{{Scally} \& {Clarke}(2002)}]{scal02}
{Scally} A., {Clarke} C., 2002, \mnras, 334, 156

\bibitem[{{Schmitt}(1997)}]{schm97}
{Schmitt} J.~H.~M.~M., 1997, \aap, 318, 215

\bibitem[{{Schneider} {et~al}\mbox{.}(2006){Schneider}, {Bontemps}, {Simon},
  {Jakob}, {Motte}, {Miller}, {Kramer}, \& {Stutzki}}]{schn06}
{Schneider} N., {Bontemps} S., {Simon} R., {Jakob} H., {Motte} F., {Miller} M.,
  {Kramer} C., {Stutzki} J., 2006, \aap, 458, 855

\bibitem[{{Schulte}(1956)}]{schu56}
{Schulte} D.~H., 1956, \apj, 124, 530

\bibitem[{{Schwarz}(1978)}]{schw78}
{Schwarz} G., 1978, The Annals of Statistics, 6, 461

\bibitem[{{Siess} {et~al}\mbox{.}(2000){Siess}, {Dufour}, \&
  {Forestini}}]{sies00}
{Siess} L., {Dufour} E., {Forestini} M., 2000, \aap, 358, 593

\bibitem[{{Skrutskie} {et~al}\mbox{.}(2006){Skrutskie}, {Cutri}, {Stiening},
  {Weinberg}, {Schneider}, {Carpenter}, \& {Beichman}}]{skru06}
{Skrutskie} M.~F., {Cutri} R.~M., {Stiening} R., {Weinberg} M.~D., {Schneider}
  S., {Carpenter} J.~M., {Beichman} C., 2006, \aj, 131, 1163

\bibitem[{{Spera} {et~al}\mbox{.}(2016){Spera}, {Mapelli}, \&
  {Jeffries}}]{sper16}
{Spera} M., {Mapelli} M., {Jeffries} R.~D., 2016, ArXiv e-prints

\bibitem[{{Spitzer}(1987)}]{spit87}
{Spitzer} L., 1987, {Dynamical evolution of globular clusters}. {Princeton
  University Press}

\bibitem[{{Spitzer}(1958)}]{spit58}
{Spitzer}, Jr. L., 1958, \apj, 127, 17

\bibitem[{{Spitzer}(1969)}]{spit69}
{Spitzer}, Jr. L., 1969, \apjl, 158, L139

\bibitem[{{Telleschi} {et~al}\mbox{.}(2005){Telleschi}, {G{\"u}del}, {Briggs},
  {Audard}, {Ness}, \& {Skinner}}]{tell05}
{Telleschi} A., {G{\"u}del} M., {Briggs} K., {Audard} M., {Ness} J., {Skinner}
  S.~L., 2005, \apj, 622, 653

\bibitem[{{Testi} {et~al}\mbox{.}(2000){Testi}, {Sargent}, {Olmi}, \&
  {Onello}}]{test00}
{Testi} L., {Sargent} A.~I., {Olmi} L., {Onello} J.~S., 2000, \apjl, 540, L53

\bibitem[{{Tobin} {et~al}\mbox{.}(2015){Tobin}, {Hartmann}, {F{\H u}r{\'e}sz},
  {Hsu}, \& {Mateo}}]{tobi15}
{Tobin} J.~J., {Hartmann} L., {F{\H u}r{\'e}sz} G., {Hsu} W.-H., {Mateo} M.,
  2015, \aj, 149, 119

\bibitem[{{Trenti} \& {van der Marel}(2013)}]{tren13}
{Trenti} M., {van der Marel} R., 2013, \mnras, 435, 3272

\bibitem[{{Upton} \& {Fingleton}(1985)}]{upto85}
{Upton} G.~J., {Fingleton} B., 1985, {Spatial data analysis by example, volume
  1: Point pattern and quantitative data.} Wiley

\bibitem[{{Vandame}(2002)}]{vand02}
{Vandame} B., 2002, in Society of Photo-Optical Instrumentation Engineers
  (SPIE) Conference Series, Vol. 4847, Astronomical Data Analysis II, {Starck}
  J.-L., {Murtagh} F.~D., eds., pp. 123--134

\bibitem[{{Vink} {et~al}\mbox{.}(2008){Vink}, {Drew}, {Steeghs}, {Wright},
  {Martin}, {G{\"a}nsicke}, {Greimel}, \& {Drake}}]{vink08}
{Vink} J.~S., {Drew} J.~E., {Steeghs} D., {Wright} N.~J., {Martin} E.~L.,
  {G{\"a}nsicke} B.~T., {Greimel} R., {Drake} J., 2008, \mnras, 387, 308

\bibitem[{{Weidner} \& {Kroupa}(2004)}]{weid04}
{Weidner} C., {Kroupa} P., 2004, \mnras, 348, 187

\bibitem[{{Weidner} \& {Kroupa}(2006)}]{weid06}
{Weidner} C., {Kroupa} P., 2006, \mnras, 365, 1333

\bibitem[{{Wolfe} {et~al}\mbox{.}(2000){Wolfe}, {Armandroff}, {Blouke},
  {Rector}, {Reed}, {Saha}, {Schommer}, {Smith}, {Smith}, \& {Walker}}]{wolf00}
{Wolfe} T. {et~al.}, 2000, in Society of Photo-Optical Instrumentation
  Engineers (SPIE) Conference Series, Vol. 3965, Sensors and Camera Systems for
  Scientific, Industrial, and Digital Photography Applications, {Blouke} M.~M.,
  {Sampat} N., {Williams} G.~M., {Yeh} T., eds., pp. 80--91

\bibitem[{{Wright} \& {Drake}(2009)}]{wrig09a}
{Wright} N.~J., {Drake} J.~J., 2009, \apjs, 184, 84

\bibitem[{{Wright} {et~al}\mbox{.}(2010{\natexlab{a}}){Wright}, {Drake}, \&
  {Civano}}]{wrig10b}
{Wright} N.~J., {Drake} J.~J., {Civano} F., 2010{\natexlab{a}}, \apj, 725, 480

\bibitem[{{Wright} {et~al}\mbox{.}(2010{\natexlab{b}}){Wright}, {Drake},
  {Drew}, \& {Vink}}]{wrig10a}
{Wright} N.~J., {Drake} J.~J., {Drew} J.~E., {Vink} J.~S., 2010{\natexlab{b}},
  \apj, 713, 871

\bibitem[{{Wright} {et~al}\mbox{.}(2014{\natexlab{a}}){Wright}, {Drake},
  {Guarcello}, {Aldcroft}, {Kashyap}, {Damiani}, {DePasquale}, \&
  {Fruscione}}]{wrig14c}
{Wright} N.~J., {Drake} J.~J., {Guarcello} M.~G., {Aldcroft} T.~L., {Kashyap}
  V.~L., {Damiani} F., {DePasquale} J., {Fruscione} A., 2014{\natexlab{a}},
  ArXiv e-prints 1408.6579

\bibitem[{{Wright} {et~al}\mbox{.}(2011){Wright}, {Drake}, {Mamajek}, \&
  {Henry}}]{wrig11b}
{Wright} N.~J., {Drake} J.~J., {Mamajek} E.~E., {Henry} G.~W., 2011, \apj, 743,
  48

\bibitem[{{Wright} {et~al}\mbox{.}(2015){Wright}, {Drew}, \&
  {Mohr-Smith}}]{wrig15a}
{Wright} N.~J., {Drew} J.~E., {Mohr-Smith} M., 2015, \mnras, 449, 741

\bibitem[{{Wright} {et~al}\mbox{.}(2014{\natexlab{b}}){Wright}, {Parker},
  {Goodwin}, \& {Drake}}]{wrig14b}
{Wright} N.~J., {Parker} R.~J., {Goodwin} S.~P., {Drake} J.~J.,
  2014{\natexlab{b}}, \mnras, 438, 639

\bibitem[{{York} {et~al}\mbox{.}(2000){York}, {Adelman}, {Anderson},
  {Anderson}, {Annis}, {Bahcall}, {Bakken}, {Barkhouser}, {Bastian}, {Berman},
  {Boroski}, {Bracker}, {Briegel}, {Briggs}, {Brinkmann}, {Brunner}, {Burles},
  {Carey}, {Carr}, {Castander}, {Chen}, {Colestock}, {Connolly}, {Crocker},
  {Csabai}, {Czarapata}, {Davis}, {Doi}, {Dombeck}, {Eisenstein}, {Ellman},
  {Elms}, {Evans}, {Fan}, {Federwitz}, {Fiscelli}, {Friedman}, {Frieman},
  {Fukugita}, {Gillespie}, {Gunn}, {Gurbani}, {de Haas}, {Haldeman}, {Harris},
  {Hayes}, {Heckman}, {Hennessy}, {Hindsley}, {Holm}, {Holmgren}, {Huang},
  {Hull}, {Husby}, {Ichikawa}, {Ichikawa}, {Ivezi{\'c}}, {Kent}, {Kim},
  {Kinney}, {Klaene}, {Kleinman}, {Kleinman}, {Knapp}, {Korienek}, {Kron},
  {Kunszt}, {Lamb}, {Lee}, {Leger}, {Limmongkol}, {Lindenmeyer}, {Long},
  {Loomis}, {Loveday}, {Lucinio}, {Lupton}, {MacKinnon}, {Mannery}, {Mantsch},
  {Margon}, {McGehee}, {McKay}, {Meiksin}, {Merelli}, {Monet}, {Munn},
  {Narayanan}, {Nash}, {Neilsen}, {Neswold}, {Newberg}, {Nichol}, {Nicinski},
  {Nonino}, {Okada}, {Okamura}, {Ostriker}, {Owen}, {Pauls}, {Peoples},
  {Peterson}, {Petravick}, {Pier}, {Pope}, {Pordes}, {Prosapio},
  {Rechenmacher}, {Quinn}, {Richards}, {Richmond}, {Rivetta}, {Rockosi},
  {Ruthmansdorfer}, {Sandford}, {Schlegel}, {Schneider}, {Sekiguchi}, {Sergey},
  {Shimasaku}, {Siegmund}, {Smee}, {Smith}, {Snedden}, {Stone}, {Stoughton},
  {Strauss}, {Stubbs}, {SubbaRao}, {Szalay}, {Szapudi}, {Szokoly}, {Thakar},
  {Tremonti}, {Tucker}, {Uomoto}, {Vanden Berk}, {Vogeley}, {Waddell}, {Wang},
  {Watanabe}, {Weinberg}, {Yanny}, \& {Yasuda}}]{york00}
{York} D.~G. {et~al.}, 2000, \aj, 120, 1579

\bibitem[{{Zinnecker}(1982)}]{zinn82}
{Zinnecker} H., 1982, Annals of the New York Academy of Sciences, 395, 226

\end{thebibliography}
\bsp

\end{document}